# The Past, Present and Future of the Resonant-Mass Gravitational Wave Detectors

**Odylio Denys Aguiar**

*Instituto Nacional de Pesquisas Espaciais, Divisão de Astrofísica, Av. dos Astronautas 1758, São José dos Campos, 12227-010 SP, Brazil, e-mail: odylio@das.inpe.br*

Running Title: **Gravitational Wave Detectors**

**Abstract.** Resonant-mass gravitational waves detectors are reviewed from the concept of gravitational waves and its mathematical derivation, using Einstein's general relativity, to the present status of bars and spherical detectors, and their prospects for the future, which include dual detectors and spheres with non-resonant transducers. The review covers not only the technical aspects of detectors and the science that will be done, but also analyses the subject in a historic perspective, covering the various detection efforts over four decades, starting from Weber's pioneering work.

## Introduction

The concept of gravitational waves goes back no more than three centuries in the history of mankind. Even after the publication of "Principia", Newton refused to commit himself as to how gravitation was transmitted from body to body across the void. His words were: "I make no hypotheses." Other scientists at that time, however, pictured gravitation as making its way through the *ether* as sound does through air [1]. It is likely that the concept of gravitational field propagation had to wait around one century before evolving to the concept of gravitational waves. It seems that only after the experimental confirmation of the existence of electromagnetic waves by H. Hertz in 1887, speculation was made about the possible existence of gravitational waves (GWs) (not only involving the gravitational field, but also the gravitomagnetic one). Heaviside in 1893, Lorentz in 1900, and Poincaré in 1905 were examples of scientists making such speculations [2]. However, a mathematical derivation of gravitational waves was only possible after the formulation of the theory of General Relativity. Einstein himself derived them, using a weak field approximation, as a irradiative solution of the vacuum equations of general relativity in 1916 [3], and in a more detailed work two years later [4]. A modified version of this derivation [5, 6] is summarized in the following section.

## 2. General Relativity and Gravitational Waves

The Einstein field equations are (neglecting the cosmological constant):

$$R_{\mu\nu} - \frac{1}{2} g_{\mu\nu} R = 8\pi \frac{G}{c^4} T_{\mu\nu} \qquad (2.1)$$

where c=2.99792458 x $10^8$ m sec$^{-1}$, and G=6.67259(85) x $10^{-11}$ m$^3$ kg$^{-1}$ sec$^{-1}$.

The left-hand side of these equations involves second order partial differential operations for the calculation of the components of the Riemann tensor. Because these equations are non-linear, the computation in GR theory is sometimes very complicated. This non-linearity of gravity also implies that the principle of superposition is no longer valid. Nevertheless, when the gravitational field is weak sufficiently far from its source many linearizations can be assumed. The spacetime curvature is nearly flat and the metric can be written as

$$g_{\mu\nu} = \eta_{\mu\nu} + h_{\mu\nu} \qquad (2.2)$$

where $\eta_{\mu\nu}$ is the Minkowski metric tensor and $h_{\mu\nu} \ll 1$. This simplifies the calculation of the Riemann tensor, which now can be expressed as:

$$R_{\alpha\beta\mu\nu} = \frac{1}{2}(h_{\alpha\nu,\beta\mu} + h_{\beta\mu,\nu\alpha} - h_{\beta\nu,\alpha\mu} - h_{\alpha\mu,\beta\nu}) \qquad (2.3)$$

where $h_{\alpha\beta,\mu\nu} = \frac{\partial^2 h_{\alpha\beta}}{\partial x^\mu \partial x^\nu}$ if terms of order $h^2$ are ignored. The Ricci and scalar tensors can now be computed and substituted into the field equations.

Because the field equations contain the $g_{\mu\nu}/2$ term, we simplify them if we define

$$\bar{h}_{\mu\nu} \equiv h_{\mu\nu} - \frac{1}{2}\eta_{\mu\nu} h \qquad (2.4)$$

where $h = h_\alpha{}^\alpha = \eta^{\alpha\beta} h_{\alpha\beta}$.

In addition to this, in order to obtain an even more compact form for the field equations, we choose a convenient gauge (the "Lorentz gauge") in which:

$$\bar{h}^{\mu\alpha}{}_{,\alpha} = 0 \qquad (2.5)$$

The field equations then assume the form

$$\Box \bar{h}_{\mu\nu} \equiv \left(-\frac{1}{c^2}\frac{\partial^2}{\partial t^2} + \nabla^2\right)\bar{h}_{\mu\nu} \equiv \bar{h}_{\mu\nu,\alpha\beta}\eta^{\alpha\beta} = -16\pi \frac{G}{c^4} T_{\mu\nu} \qquad (2.6)$$

In vacuum $T_{\mu\nu} = 0$, and we have a tensor wave equation with solutions of the form

$$\bar{h}_{\mu\nu} = A_{\mu\nu} e^{[ik(z-ct)]} \qquad (2.7)$$

which represents a monochromatic wave of spacetime geometry [7] propagating along the +z direction with speed c (other theories of gravitational might predict a speed different from that of light) and frequency kc. These waves, ripples in the curvature of spacetime, are so called gravitational waves.

In an analogy with electromagnetic waves, which can be generated by accelerated charges, gravitational waves are produced by accelerated masses. But there are some major differences between the two. The conservation of momentum, plus the fact that mass comes with only one sign, positive, prevents an oscillating (or accelerating) mass dipole from radiating gravitational waves [8]. Only accelerated mass quadrupole or higher multipole moments can produce gravitational waves. Another difference is that the gravitational waves are very weak compared to the electromagnetic waves. This is a natural consequence of the weakness of the gravitational forces compared to the electromagnetic force (36 orders of magnitude weaker for elementary particles such as the proton).

$$\frac{G\, m_p^2}{\frac{1}{4\pi\varepsilon_o} e^2} \sim 10^{-36}, \text{ where } m_p \text{ is the proton mass and } e \text{ its charge}.$$

This fact added to the impossibility of the emission of dipole gravitational radiation, makes the generation of gravitational waves in nature only potentially significant for a very large total amount of mass in a coherent, fast, and strongly accelerating movement of particles. Therefore, only astrophysical and cosmological events are potentially detectable for the first set of successful gravitational wave detectors.

This weak coupling between mass and gravitational waves is reciprocal. Gravitational waves, as it would be expected, also couple very weakly with matter. That is why the gravitons decoupled from matter when the Universe was only ~$10^{-43}$ s old (the Planck time). In one sense this is good: gravitational waves can bring information from the dense cores of some astrophysical objects without being significantly attenuated by crossing the outside layers of matter, something electromagnetic waves are unable to do. But on the other hand it is bad: this weak coupling of GWs with matter makes it extremely difficult for the experimentalist to detect them. Actually, since its mathematical prediction in 1916, no single gravitational radiation pulse has ever been detected directly. The only observational confirmation of the existence of such radiation comes from the observed decrease in the orbital period of a few compact binary systems such as the PSR 1913+16 system discovered by Hulse and Taylor [9] in 1974 (some of the other ones are: PSR J0737-3039, PSR 2127+11C, PSR 1534+12, and J1141-6545). After ruling out other possible mechanisms to explain the observed change of orbital period, gravitational radiation damping is left as the only probable cause. The GR theory prediction for the orbital period decrease due to the loss of energy by gravitational waves in these systems agrees within 0.2% with observation [10]. This discovery gave to Taylor and Hulse the 1993 Nobel Prize in physics [11].

In order to understand the interaction with matter of a locally plane gravitational wave in this linearized theory, we change the gauge, once more, to one that is transverse and traceless. In this gauge only spatial components of $h_{\mu\nu}$ are nonzero ($h_{\mu 0} = 0$), and they are transverse to the direction of propagation. Furthermore, these components are divergence-free ($h_{kj,\,j} = 0$) and trace-free ($h_{kk} = 0$). And since $h = h_\alpha{}^\alpha = h_{kk} = 0$, we can conclude that

$$\bar{h}_{\mu\nu} = h_{\mu\nu} = h_{\mu\nu}^{TT}. \tag{2.8}$$

This gauge is called a TT or transverse-traceless gauge.[3] Here, the Riemann curvature tensor assumes the simple form

$$R_{j0k0} = -\frac{1}{2} h_{jk,\,00}^{TT} \tag{2.9}$$

and in particular [12]

$$R_{x0x0} = -R_{y0y0} = -\frac{1}{2}\ddot{h}_+\left(t - \frac{z}{c}\right) \quad \text{and} \quad R_{x0y0} = R_{y0x0} = -\frac{1}{2}\ddot{h}_\mathbf{x}\left(t - \frac{z}{c}\right) \tag{2.10}$$

where [13]
$$h_+ \equiv h_{xx}^{TT} = -h_{yy}^{TT} = \mathcal{R}\left\{A_+ e^{-i[\omega(t-z/c)+\phi_+]}\right\} \quad (2.11)$$
and
$$h_{\mathbf{x}} \equiv h_{xy}^{TT} = h_{yx}^{TT} = \mathcal{R}\left\{A_{\mathbf{x}} e^{-i[\omega(t-z/c)+\phi_{\mathbf{x}}]}\right\} \quad (2.12)$$
and $A_+$ together with $A_{\mathbf{x}}$ are the amplitudes of the two independent modes of polarization (+ and x ). Then, the gravitational wave driving force acting on each element of mass $m_i$ of a material body can be derived as
$$F_j = \frac{1}{2} m_i \ddot{h}_{jk}^{TT} k, \quad \text{where } j, k = x \text{ or } y, \quad (2.13)$$
and the total force ($= F_x \mathbf{e}_x + F_y \mathbf{e}_y$) becomes
$$\vec{F} = \frac{1}{2} m_i \left\{ \left( \ddot{h}_+ x + \ddot{h}_{\mathbf{x}} y \right) \mathbf{e}_x + \left( \ddot{h}_{\mathbf{x}} x - \ddot{h}_+ y \right) \mathbf{e}_y \right\}. \quad (2.14)$$

From equations 2.11 and 2.12: if $\phi_+ = \phi_{\mathbf{x}}$, the resultant force is linearly polarized; if $\phi_+ = \phi_{\mathbf{x}} \pm 90°$ and the magnitudes $A_+$ and $A_{\mathbf{x}}$ are equal, the resultant force is circularly polarized; otherwise it is elliptically polarized.

Forces induced by gravitational waves are strictly transverse, as in Fig. 2.1, which shows the effect of pure $h_+$ or $h_{\mathbf{x}}$ plane gravitational waves arriving perpendicularly to the plane of a square array of test particles (the amplitudes are exaggerated). It is apparent that GWs not only change distances, but also angles. The orthogonality between $h_+$ or $h_{\mathbf{x}}$ can be appreciated if we imagine deformations for very small amplitudes; a material segment forming an angle $\phi = 0°, 90°, 180°,$ and $270°$ with the x axis will change length due to $h_+$ and rotate due to $h_{\mathbf{x}}$ and vice versa if $\phi = 45°, 135°, 225°,$ and $315°$. It is the Principle of Equivalence that causes the acceleration (and the force) to be locally undetectable ($\xi_k = 0$). Only the relative acceleration or the "relative force" can be observed.

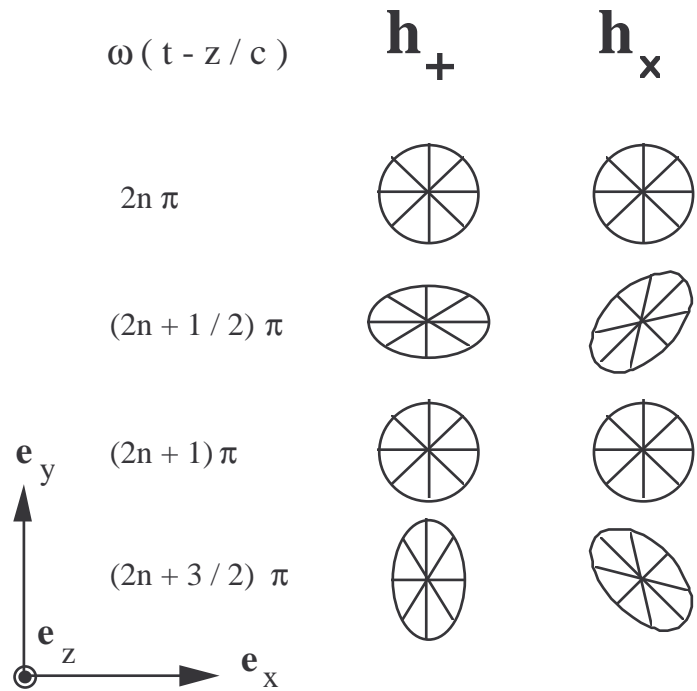

**Figure 2.1.** Deformation of a circle due to forces induced by pure $h_+$ or $h_{\mathbf{x}}$ waves.

The strain, $h = \sqrt{h_+^2 + h_\times^2}$, $h \equiv \Delta L / L$, is the dimensionless amplitude that can be measured by gravitational wave detectors. It is the composition of the two dimensionless polarization amplitudes $h_+$ and $h_\times$. However, a more useful quantity, which gives a better indication of the detector's sensitivity, is the "strain spectral sensitivity", in units of Hz $^{-1/2}$. This quantity takes into account the observable frequency bandwidth where the signal is present. Evidently, this means that burst events should be strong in order to be detected, in contrast with detectable monochromatic signals, which can be much fainter.

It is possible to envision various schemes of detection of gravitational waves [5, 14]. The two mostly explored principles of detection are the monitoring of the distance between two or more points and the measurement of the oscillations driven on a solid "antenna" by stress [15]. The first principle of detection can be performed with a laser interferometer, or a doppler tracking array [16, 17]. The second principle can be carried out with a resonant-mass antenna coupled to an electromechanical transducer of some kind. Joseph Weber was the first one to propose the feasibility of this.

## 3. The First Generation of Resonant-mass Gravitational Wave Detectors: the Pioneering Work of Joseph Weber and many others

By the end of the 50's, after four decades of debate, it became possible "for a number of physicists to conclude that general relativity really does predict the existence of gravitational waves". Also, thanks to the technological advances in the 40's and 50's, it became feasible to do some new gravitational experiments and to repeat the older ones in more precise ways. The man who was capable of making these two statements at that time was Joseph Weber [6]. He was, at the same time, a brilliant experimentalist and a competent theorist on general relativity. He almost invented the maser in the earlier 50's. On the other hand, he spent the 1955-56 academic year as a fellow of the Institute for Advanced Study in Princeton, New Jersey, where he dipped himself in general relativity. This rare union between experiment and theory made him the only one with "vision to see that there were technological possibilities of ultimate success" [13]. His major contribution to science was to start this experimental and exciting field of gravitational wave detection, by constructing the first resonant-mass antenna with eletromechanical transducers and influencing Robert Forward (his former Ph.D. thesis student) to construct the first free-mass antenna with laser interferometry [18]. The idea of a laser-interferometer gravitational-wave detector was independently proposed by Felix Pirani in 1956, Gertsenshtein and Pustovoid in 1962, Weber in 1964, and Rainer Weiss [13]. Let us now understand these two techniques.

The Equivalence Principle forbids that the accelerations produced by the passage of gravitational waves be felt locally. Any detector has to be an extensive one, because the accelerations are of the tidal type. This poses two possibilities: a free-mass detector or a non-free-mass detector. In the free-mass case, the experimentalist should monitor the distance between the two or more free-masses that compose the detector. No energy is absorbed from the gravitational wave in this case. Weber proposed to use laser interferometry to monitor the distances between these masses (actually this becomes an almost-free-mass case, because on Earth it is impossible to have a completely free-mass situation). A detailed explanation of this technique can be found in another chapter of this book.

In the case of non-free-mass, there is some sort of connection between the masses, such as the forces that connect the atoms in the crystalline structure of a body. This is the case of the so called resonant-mass detector. The atoms of a body try to follow the geodesic trajectories produced by the spacetime distortion (geodesic deviation) caused by the passage of the gravitational wave. However, the electrostatic connections between the body atoms prevent them from following these precise trajectories. This case is analyzed below.

Suppose we model the mechanical longitudinal oscillation of a cylindrical bar antenna by the oscillation of two point masses separated by the distance r and connected by a spring. For simplification, let us consider the detector perpendicular to the wave propagation direction, so that we can write the equation that describes the oscillation of the point masses as:

$$\ddot{\zeta} + \frac{b}{m}\dot{\zeta} + \frac{k}{m}\zeta = \frac{1}{2}r\ddot{h}(t) + F_N \tag{3.1}$$

(acceleration + dissipation force + restoring force = external force caused by the GW + external force caused by various noise sources), where $f_o$ is the resonant frequency, $\omega_o = 2\pi f_o$, b is the dissipation factor, m is the mass of each point mass, k is the spring constant, $\omega_o^2 = \frac{k}{m}$,

$Q = \frac{\omega_o m}{b}$ (the mechanical quality factor),

$\ddot{h}(t) = \sqrt{[\ddot{h}_+(t)]^2 + [\ddot{h}_x(t)]^2}$,

$\ddot{h}_+(t) = -2c^2 R_{x0x0} = 2c^2 R_{y0y0}$, and $\ddot{h}_x(t) = -2c^2 R_{x0y0} = -2c^2 R_{y0x0}$

The Riemann tensor is the driving force in this harmonic oscillator equation.

Making h (t) = $h_o$ sin $\omega t$, so that $\ddot{h}$ (t) = - $h_o$ $\omega^2$ sin $\omega t$, we arrive to the following results [19]:

$$|\zeta(t_p)| \sim \frac{1}{4}rh_o\omega_o t_p \quad \text{if } t_p << Q/\omega_o \tag{3.2}$$

and

$$|\zeta(t_p)| \sim \frac{1}{2}rh_o Q \quad \text{if } t_p \to \infty, \tag{3.3}$$

where $t_p$ is the time duration of the pulse.

The problem with this scheme was noise, more specifically: thermal noise. The average amplitude of oscillation of a cylindrical bar due to thermal fluctuations (Brownian-noise forces or Langevin forces) was many orders of magnitude larger than the oscillation amplitudes induced by an expected gravitational wave. Weber envisaged a very clever solution to this problem. He had the idea of using a material with high mechanical quality factor (Q = ω $\tau_E$, where $\tau_E$ is the energy relaxation time, $\tau_E$ = (maximum stored energy)/(power dissipated); $\tau_A$ , the amplitude decay time is 2 $\tau_E$). If one uses high Q materials for the antenna, the variations of amplitude due to thermal fluctuations will still be very large (the same as before), but they would occur in a time much larger than the gravitational wave period, so large, that the gravitational wave signal would still be detectable if a suitable digital filter is used [20, 5].

Weber's first large suspended bar antenna of high Q aluminum was constructed in the early 60's. It was a 1.2-ton aluminum cylinder of length ~ 1.5 m and diameter of ~ 61 cm suspended in a vacuum chamber on acoustic filters. Its first longitudinal mode was resonant at room temperature at 1657 Hz [21]. The transducer system, able to convert the mechanical strains to voltages, was a series of piezoelectric crystals (quartz strain gauges) mounted on its surface close to the central region as shown in figure 3.1a. It started operating with good sensitivity and isolation around January 1965. Two years later, Weber reported the first possible gravitational wave signals [22].

In 1968, he again reported possible events. This time, he used two aluminum cylinders tuned to ~ 1.66 kHz instead of just one. They were spaced about 2 km apart. He claimed that the number of coincident events had an extremely small probability to be statistical [23]. Finally, in 1969, Weber stated that coincidences had been observed on gravitational-radiation detectors over a base line of about 1000 km at Argonne National Laboratory and at the University of Maryland, and that the probability that all of these coincidences were accidental was incredibly small [24]. The output of each detector was put on a chart recorder like those used to record earthquakes. By 1973, he had claimed an excess of coincidences from statistical average of about seven events per day peaked in the direction of the galactic center [25, 26, 27]. These and subsequent observations by Weber were greeted with great excitement in the early 1970s; however, the strength implied by his signals was very much in excess of what was expected.

In the following years, various experimenters built more sensitive bars, including low-temperature bars, and looked for signals, but none of them confirmed Weber's findings. The first ones to find null results were Tyson at Bell Labs, Murray Hill, NJ, and Levine and Garwin at the IBM Thomas J. Watson Research Center, in Yorktown Heights, NY, both in 1973. [28, 29, 30]. Both these experiments disagreed with Weber's results and suggested that the events he found were not gravitational wave events.

In any case, Weber's pioneering work was decisive for the initial growth of the gravitational wave community. Thanks to Weber's reported results in 1969 of having seen GW signals, about ten groups tried to repeat his results. Even though none could confirm Weber's findings, the basis of gravitational wave detection was definitely established.

The following 12 groups, which operated room temperature resonant-mass GW detectors in the 60's, 70's and 80's, were motivated by Weber's pioneering effort:
- <u>Moscow, in Russia</u>: *two detectors 20 km apart composed of two 1.2 ton aluminum alloy bars*, 150 cm long by ~ 60 cm diameter, resonant at 1640 Hz, and equipped with capacitive transducers [31];

- <u>BTL (Bell Labs), New Jersey, USA</u>: *one 3.7 ton aluminum alloy bar*, 357 cm long, ~70 cm diameter, resonant at 710 Hz, equipped with PZT-8 transducers [28];
- <u>Rochester, in Rochester (NY), USA</u>: *another 3.7 ton aluminum alloy bar 420 km apart from the one above,* 357 cm long, ~70 cm diameter, resonant at 710 Hz, equipped with PZT-8 transducers [32];
- <u>IBM, Yorktown Heights (NY), USA</u>: *one 118 kg aluminum alloy bar*, 150 cm long by 19 cm diameter, resonant at 1695 Hz, equipped with PZT-4 transducers [29, 30];
- <u>Bristol group, England</u>: *two parallel split-bars in the same vacuum chamber composed of two aluminum half-bars.* Each split-bar has its own transducer (made of lithium niobate piezoelectric material) and amplifier detector, but they share a common vacuum chamber and vibration isolation system. The signals from the two split-bars are monitored separately and also cross-correlated [33];
- <u>Glasgow, Scotland</u>: *two split-bar detectors 50 m apart composed of two aluminum half-bars of 300 kg total mass and 155 cm total length*, ~30 cm diameter, cemented together via PZT transducers, resonant at 1020 and 1100 Hz [34];
- <u>Reading-Rutherford Lab, England</u>: *two split-bar detectors 30 km apart composed of two aluminum alloy half-bars of 625 kg total mass and 150 cm total length*, 46 cm diameter, cemented together via PZT transducers, resonant at 1200 Hz [35];
- <u>Univ. Tokyo, Tokyo, Japan</u>: *two 1.4 ton 165cmx165cmx19cm square antennae made of an aluminum alloy*, both in the Physics building, resonant at 145 Hz, equipped with dc-capacitive transducers [36] (later, one of these antennae was mechanically tuned to 60.2 Hz in order to become CRAB II [37]), $h < 8.4 \times 10^{-21}$ for continuous waves, *one 400 kg 110cmx110cmx12cm square antenna made of an aluminum alloy*, resonant at 60.2 Hz, equipped with dc-capacitive transducers (CRAB I) [38], $h < 1.1 \times 10^{-19}$ for continuous waves, and some other small antennae (M < 40 kg) [39, 40];
- <u>Munich-Frascati group</u>: *two detectors ~700 km apart (later only 10 km, when the Frascati detector was moved to Garching) composed of two 1.2 ton 6061-O aluminum alloy bars*, 154 cm long by 62.5 cm diameter, resonant at 1654 Hz, equipped with piezoelectric transducers, and intended to reproduce with precision the Weber experiment, with some sensitivity improvements [41, 42, 43], set the lowest upper limits to the rates of gravitational wave pulses in the 70's [44]. They took a piezoelectric material with better mechanical and electrical properties, and arranged the piezos differently (arranging strain and polarization parallel) thereby providing a better coupling. Because of the new topological arrangement of the piezos, they were able to manage an impedance matching between the "source" of the signal, the piezos, and the input of the amplifier. Thus, the detector was noticeably improved in comparison with Weber's setup. In addition, the signal processing was performed with the two degrees of freedom in phase space, equivalent to amplitude and phase or the two independent quadrants, whereas in most cases Weber used just the energy of the bar [45];
- <u>Zhongshan Univ., Guangzhou, China</u>: *one ~2 ton(1963 kg) aluminum alloy bar*, 178 cm long, 71.4 cm diam., resonant at ~1.5 kHz, using PZT-4 transducers, and *one 498 kg 200cm$^2$ square antenna*, resonant at 47.3 Hz, equipped with dc-capacitive transducers [46, 47];
- <u>Beijing Univ., Beijing, China</u>: *one 1.3 ton aluminum alloy bar*, 153 cm long, 62.5 cm diameter, resonant at 1687 kHz, equipped with PZT transducers [48, 49];
- <u>Meudon group, in France</u> : *a conical antenna equipped with capacitive transducers* [50].

Also six groups, Stanford and LSU in the USA, University of Rome, University of Western Australia (UWA) [51], University of Regina (Canada) [52, 53], and the

Legnaro group, were formed in the 60's, 70's and 80's. They decided to construct cryogenic resonant-mass (RM) GW detectors, instead of room temperature ones, starting a second and a third generation of RM detectors. At the same time some other existing groups, such as the ones in Tokyo (this in collaboration with KEK), Moscow [54], Rochester [55], and Maryland [56] were switching to cryogenics. "These groups made a number of significant improvements over Weber's original design. One improvement was to lower the temperature of the bar to liquid helium temperatures (4 Kelvin). The second was a better suspension of the bar with increased vibration isolation. A third was the use of a resonant transducer and low noise amplifier to observe the motion of the bar." [57]. Figure 3.1 shows some historical pictures [58] and author's private archive).

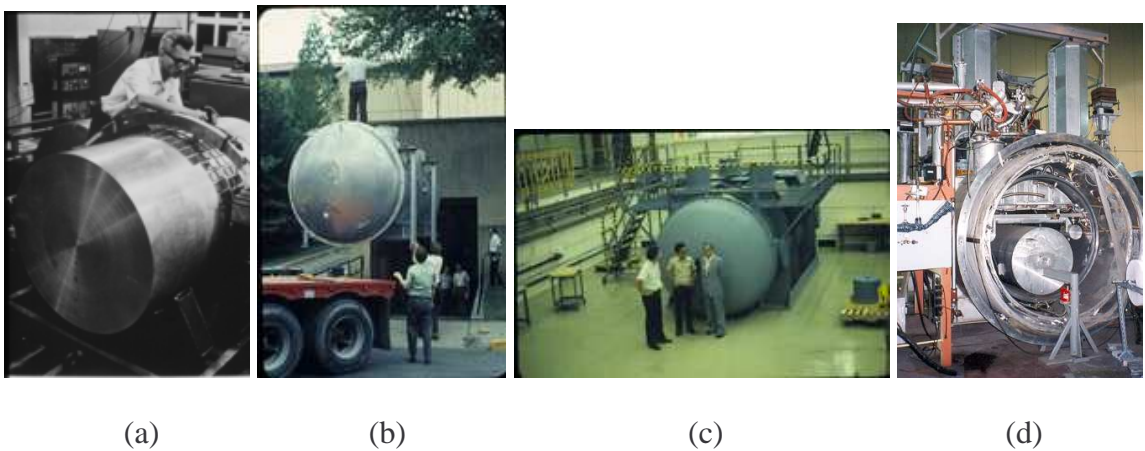

(a)          (b)          (c)          (d)

**Figure 3.1.** a) Weber and one of his room temperature bars (60's) (courtesy of the LSU Gravity Allegro group); b) the LSU detector dewar arriving to the old lab (early 70's) (courtesy of the LSU Gravity Allegro group); c) the Explorer detector at CERN (early 80's) (courtesy of the LSU Gravity Allegro group); d) the earthquake damaged Stanford bar detector (picture taken in 1994) with the large 4.8 ton 6061 aluminum alloy antenna inside (author's private collection).

Let us finish this section by mentioning the standard data analysis procedure used by the first and second generation detectors. This procedure was to produce a list of events above some arbitrary reasonable energy threshold for each detector and from these lists to produce time delay histograms for each pair combination of detectors. The time delay histograms would give the number of coincidences for each combination of time shift (delay or advance) of their detector event list time. The "zero time delay bin" would, then, give the number of coincidences between the two detectors without altering their event list times. Evidently, if no gravitational wave detection was made all the histogram bins (including the zero delay bin) would have their counts fluctuating around an average number, which was related to the statistical probability of coincidence by chance. However, if some GW detections were made the "zero time delay bin" would have additional counts. Consequently, if the "zero time delay bin" was significantly (statistically) above the average count, as Weber found many times in the late 60's and 70's, a suspicion of detection could be claimed.

In the 90's the cryogenic bar detectors groups decided to make some changes in the way they were producing their event lists and the way they were comparing them with each other. They decided to improve the statistical results of this kind of analysis, decreasing the number of events in their detector lists by increasing the energy threshold. This threshold was chose to produce only about 100 events/hour for each detector. With this new procedure the average number of coincidences by chance would drop and, consequently, the significance of a few real detections would increase statistically, standing out more clearly. Another procedure had to do with the way the groups exchanged data. They started to send four lists of events to the other groups in which only one of them had the correct time. The reason for that "protocol" was to avoid any analysis bias.

Evidently when three or more detectors find an event at the same time a more strong case is made. Unfortunately no triple or quadruple events were ever registered for the combinations of lists (with reasonable high threshold). Part of the reason for this was because it is hard to coordinate the operation of individual bars in order to have them operating synchronously. In the 80's there was only about 60 hours of triple operation between (Stanford, LSU and Rome). In the 90's, Stanford was no longer active, Niobe only started in 1993 and Auriga in 1997. From January 1997 to June 2003 (a total of 78 months), triple operations were much more frequent, about two years (~30% of the total time), and quadruple operation were possible for four months (~5%).

## 4. The second generation resonant-mass detectors: the advantage of cooling the mass down to low temperatures

There are many advantages of cooling the antenna down to low temperatures: the thermal (Brownian) noise decreases, the antenna mechanical quality factor ($Q_{mec}$) increases (a fact discovered later) [59], superconductor materials can be used for the transducer circuitry, increasing the transducer electrical quality factor ($Q_e$), and very low noise cryogenic amplifiers can be used.

Basically, a second or third generation resonant-mass detector consists of a cryogenic resonant-mass antenna coupled to an electromechanical resonant (at the same frequency) transducer that has its electrical output signal pre-amplified by a very low noise cryogenic amplifier. This pre-amplified signal is recorded on magnetic tape, after passing an analog-digital converter, for digital filtering and computational analysis later. The cryogenic antenna which is made of a high mechanical Q material, usually an aluminum alloy (the Australian Niobe antenna is one of the exceptions, because it is made of pure niobium), is kept in vacuum and at 4.2 K, the boiling point of liquid helium (second generation) or at ~ 0.1 K if a dilution refrigerator is used (third generation), and isolated as much as possible from floor or sound vibrations. The first transducer was composed of piezoelectric crystals during the first generation (room temperature) of antennae. They were replaced around the mid-seventies by electromechanical resonant transducers using superconducting circuits. The first one was a tunable-diaphragm with an inductive pick-up transducer, designed and constructed by Ho Jung Paik [60]. Compared to capacitive or inductive resonant transducers, piezoelectrical transducers did not have high mechanical quality factors, even at low temperatures, and electromechanical couplings were not as high.

Cryogenic mass-resonant antennae of the bar type have their first longitudinal vibrational mode, which is sensitive to a quadrupolar gravitational wave, around 900

Hz. The 5056 aluminum alloy (Mg 5.1%, Mn 0.12%, Cr 0.12%, balance Al) is the one most used, because it is known by the gravitational wave community as the aluminum alloy with the highest mechanical Q at cryogenic temperatures [61].

As an example of second generation bar antenna, Fig. 4.1 shows the schematic drawing of the LSU detector [62] and a picture of the detector at the new lab [63]. The antenna is balanced on a titanium cable which is attached to the intermediate mass, an H-shape, 2.5-ton bronze casting supported by titanium rods that hang from vibration isolation stacks. These stacks are kept in vacuum chambers which are connected to the 2 K shell space. To avoid overloading the schematic, the aluminum frame and pneumatic air springs on which the stacks rest were omitted from the drawing. The transducer is bolted to one of the bar ends, and a calibrator capacitor is bolted to the other end. All the wiring attached to the transducer passes through small intermediate brass discs connected to each other by fine titanium wires. This array of discs is named a "Taber isolation" stack after its inventor, and its function is to guarantee vibration isolation for the antenna-transducer system. The bronze X mass, the vibration isolation on top, and the "Taber isolation" stack work together to reflect the external mechanical excitation by the principle of mechanical impedance mismatching.

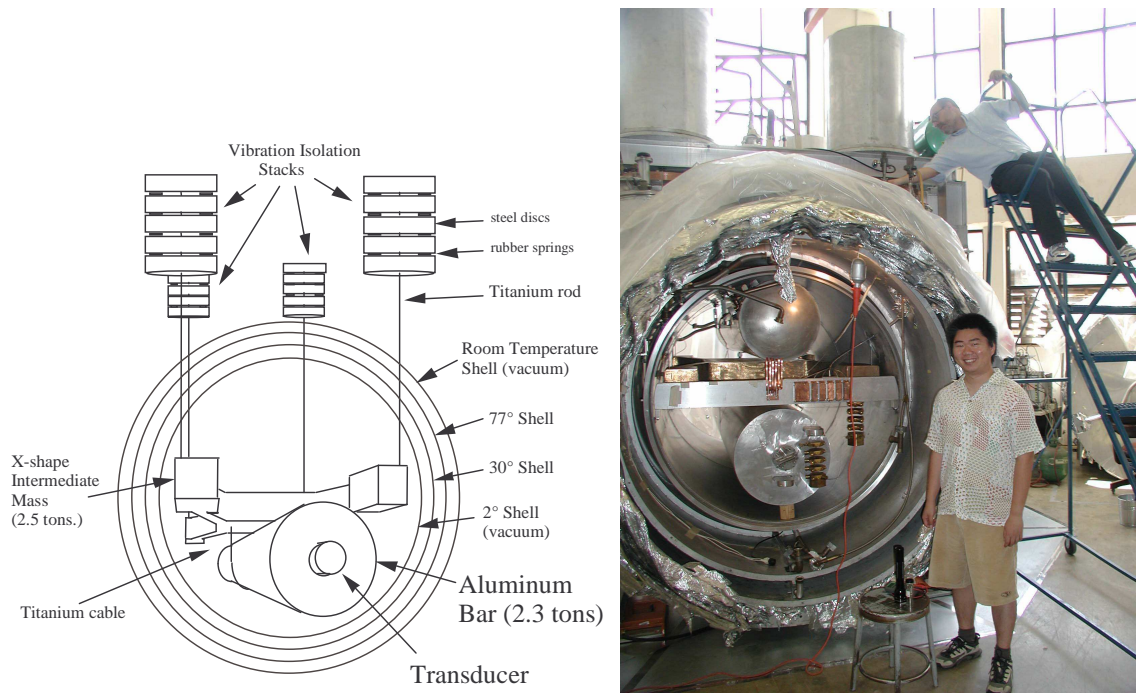

**Figure 4.1.** The LSU Cryogenic Antenna. a) Schematic view (the extra liquid helium cylindrical tank over the intermediate mass and the bar antenna was omitted for simplicity); b) Picture of the detector (without the end caps) in the new lab, with Warren Johnson (top) and Ik Siong Heng (courtesy of the LSU Gravity Allegro group).

The principle of detection is very simple. When gravitational waves strike the aluminum bar antenna, it is driven into oscillation. A major part of the energy received from the GW is coupled to the bar's first longitudinal mode which has a node about the titanium cable section and anti-nodes at both ends of the bar. The transducer attached to one of the ends is, therefore, set into oscillation. Because the transducer mechanical resonance is at the same frequency $f_o$ as the bar resonance frequency, there is a beating between the two oscillations with beat frequency equal to $f_o/2$ times the square root of the ratio: transducer effective mass / antenna effective mass. The amplitude of the transducer harmonic oscillator increases until almost all the energy from the bar's first longitudinal mode has been transferred to the transducer oscillation mode. This method obtains an amplitude gain equal to the square root of the inverse ratio mentioned above, and an increase in frequency bandwidth above noise level. The transducer then converts part of this energy into an electrical signal which is pre-amplified by a very low noise cryogenic amplifier. Finally, this amplified signal, as mentioned before, is recorded on magnetic tape for analysis later.

As we have stated in section 2, gravitational waves have a very weak interaction with matter. If we calculate the integrated cross section $\int_0^\infty \sigma_n(\nu)d\nu$ of a bar antenna of mass M, length L, radius R, and Poison ratio $\sigma$, for a sinusoidal gravitational wave with optimum polarization the result is [64]

$$\int_0^\infty \sigma_n(\nu)d\nu = \left(\frac{8}{\pi}\right)\left(\frac{M}{n^2}\right)\left(\frac{G}{c}\right)\left(\frac{V}{c}\right)^2 \sin^4\theta_i \left[1+\frac{\sigma(1-2\sigma)}{2}\left(\frac{n\pi R}{L}\right)^2\right] \quad (4.1)$$

where n is the longitudinal mode number, V is the material sound speed, and $\theta_i$ is the angle the wave propagation vector forms with the bar axis. For the first longitudinal mode (n= 1) of the LSU antenna (M=2300 kg, L=3 m, R= 0.6 m, v=5.4 km/sec, and $\sigma$=.345) and $\theta_i = 90°$, this equation gives $4 \times 10^{-25}$ m$^2$ or 4 kbarns, which is $4.5 \times 10^{24}$ times smaller than the bar's physical cross section area.

This equation tells us that we can increase the cross section if we use higher density materials or material with a higher sound velocity. If we include a transducer resonant at the same frequency as the bar antenna, the equation changes a little because the mechanical system is no longer a single harmonic oscillator, but now has two normal modes. However, the integrated cross section continues to be dependent on the first power of M and the second power of v. But a more stringent equation is obtained if we take into account the noise involved in the process. We can define a noise temperature $T_N$ which, when multiplied by the Boltzmann constant, is equal to the minimum energy deposited in the antenna by the GW pulse that can provide a signal to noise ratio equal to one after optimum filtering. The rms conventional gravitational wave strain amplitude for the bar antenna as a function of this noise temperature can be found to be, at f ~ 900 Hz for any of the current bar antenna, approximately equal to [65]

$$h = \frac{\pi}{4}\frac{1}{L}\sqrt{\frac{4 k_B T_N}{M \omega^2}} \sim 10^{-18}\sqrt{\frac{T_N}{20 \text{ mK}}} \quad (4.2)$$

This means that by decreasing the noise temperature or by increasing the mass one achieves a smaller h. More recently, resonant-mass are using sensitivities $\tilde{h}$ in units of Hz$^{-1/2}$. So, the dimensionless sensitivity h for bursts depends on $\tilde{h}$ and on the bandwidth B. If $\tilde{h}$ is constant along the bandwidth B, for SNR = 1 and a 1 ms burst, h can be found using the following expression [66]:

$$h = \frac{\tilde{h}}{0.001}\sqrt{\frac{2}{\pi B}} \quad (4.3)$$

(By the way, the expression valid for continuous monochromatic GW signals is completely different from the above one; there h becomes numerically much smaller than $\tilde{h}$, and the duration of the time the data is accumulated is taken into account).

The noise temperature could be lowered by increasing the mechanical Q of the antenna, or by cooling the antenna with a $He^3$-$He^4$ dilution refrigerator down to 0.1-0.01 K together with the use of less noisy transducers and amplifiers. A less noisy transducer can be obtained by increasing its mechanical and electrical Q's, and a less noisy amplification can be achieved by the use of a D.C. SQUID preamplifier or HEMTs amplifiers. The present state of the art SQUIDs can reach ~ 2 $\hbar$ of energy sensitivity [67] in units of J $Hz^{-1}$. The bandwidth can be expanded by the use of a transducer with more stages [68] (bandwidth = $2 f_o \sqrt{\mu}$, where $\mu$ is the ratio between two consecutive masses). The mass can be increased by a factor of 15 by the use of a spherical antenna. This also has the advantage of being omnidirectional, of measuring all the GW polarizations, and of determining the direction from which the wave arrives [69]. Noise temperatures below the "standard Heisenberg quantum limit" for a 1 kHz resonant-mass bar detector ($T_N \approx \hbar \omega / k_B$ = 0.05 μK) could be achieved by circumventing techniques that perform quantum non-demolition measurements [70].

The first to start talking about cryogenic detectors was William Fairbank from the Stanford University, around the time of Weber's first PRL paper (1966) [71]. Fairbank discussed this idea with Bill Hamilton many times over the years. However, it was only when Weber strongly claimed evidences for the discovery of gravitational radiation [24] that Fairbank proposed the construction of cryogenic detectors [72]. J. M. Reynolds a professor at LSU at that time suggested to Fairbank and Hamilton the use of a large available laboratory at the LSU physics building, initially constructed for a Van De Graaff accelerator, to house one of the bar detectors.

The experimental search for gravitational waves started in Italy at the University of Rome in September 1970 (the Frascati group of ESRIN, which took part in the Frascati-Munich collaboration does not count as an Italian experiment, but rather as a European ESRO experiment). The group was formed initially by Edoardo Amaldi, Guido Pizzella and Massimo Cerdonio. Amaldi had attended lectures by Joe Weber in 1961 in Varenna and visited him at the University of Maryland in 1966. He was enthusiastic to start this experimental field in Italy, which became possible with the manifested interest of Pizzella and Cerdonio [73].

In January 1971 Amaldi confidentially received the Stanford and Louisiana (Fairbank-Hamilton) proposal for a detector consisting of a 5 ton ultracryogenic (3mK) aluminum bar antenna, equipped with a DC SQUID amplifier, and coupled to a resonant transducer by Remo Ruffini. This fact influenced the recently formed Italian group to collaborate in this effort. Pizzella, Cerdonio, Renzo Marconero, and Ivo Modena, with the help of Ruffini, visited Stanford, LSU, NASA, and Bell Labs to start collaboration on April $19^{th}$, 1971 [73].

A collaboration then started between Louisiana, Rome and Stanford for the construction of three 5-ton cryogenic detectors to be installed in the three locations [73]. The cryostats for the Stanford and Louisiana detectors were constructed at the Mississippi-NASA Test Facility Center in Louisiana [74]. The one in Rome was designed and constructed in Italy. However, at that time the technology for welding aluminum was not well known in Europe and the construction of a reliable big cryostat could not go fast. When the Rome group realized this they decided to suspend activities with the large cryostat and to continue the experiment through successive steps with smaller cryostats [75] until the construction of ALTAIR (M = 389 kg) [76], Explorer (M = 2.3 ton) at the CERN (Figure 4.2) and the ultracryogenic (~100 mK) Nautilus (M = 2.3 ton initially, and now lighter for searching the SN1987A pulsar). The Rome group also constructed a bar, similar to Explorer, but which would operate at room temperature and was equipped with PZT and FET amplifiers. It was GEOGRAV, which was in operation by the time of SN1987A [77, 78]. They also constructed the AGATA room temperature antenna. AGATA was installed in 1980 at CERN and in the beginning it was mainly used for tests. It later gathered data for long periods [79].

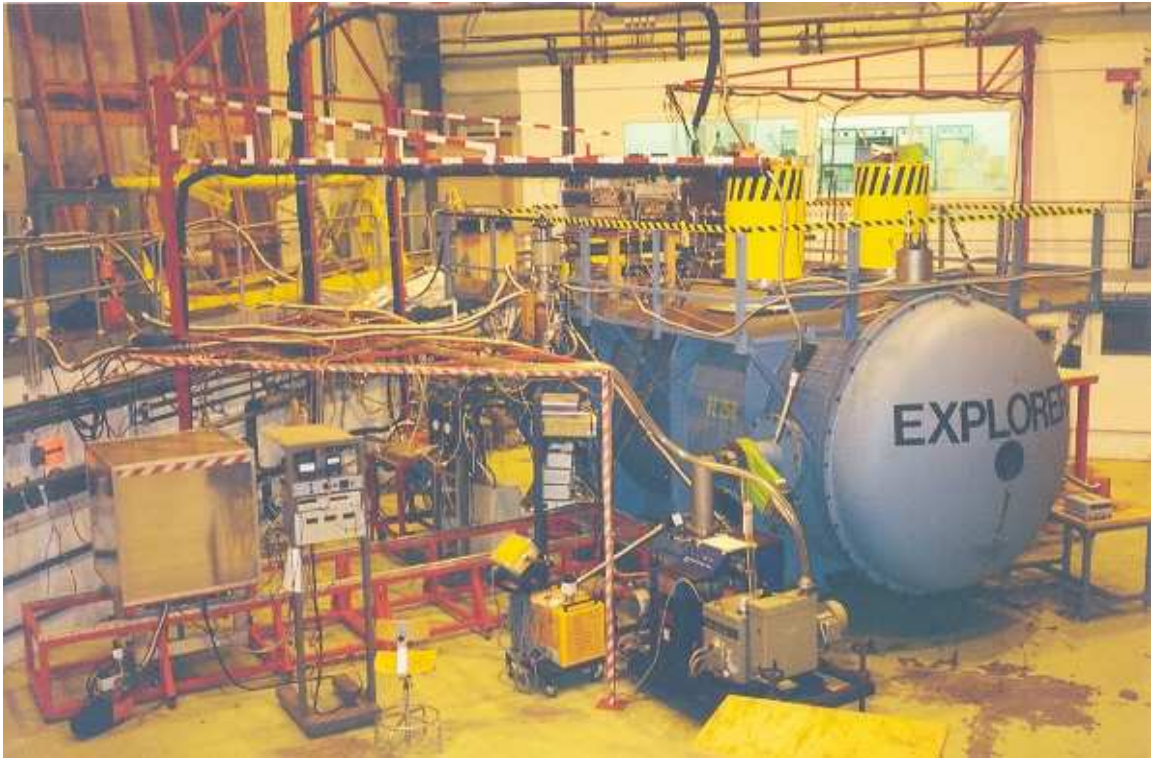

**Figure 4.2.** The Explorer detector at CERN (author's private collection).

By 1976 both the groups in Stanford and the Rome had already observed the mechanical Nyquist or Johnson (Brownian) noise at 4.2 K in their initial cryogenic gravitational-wave antennae [80, 75]. The Stanford group was using a 680 kg aluminum alloy antenna, 0.4 m in diameter, 2 m long, covered with a 0.4-mm sheet of superconducting Nb-Ti, which was levitated by a magnetic field of 0.2 T. The antenna was resonant at 1315.3 Hz and presented an effective noise temperature of 0.39 K. On the other hand, the Italians used a small 30 kg bar, equipped with a large piezoelectric crystal and a FET electronic amplifier. The noise temperature obtained was ~10 K.

The Rome group achieved an important step by cooling and measuring the Brownian noise of an intermediate size antenna of ~400 kg (ALTAIR) [76]. Noise temperatures as low as 300 mK (0.3 K) were obtained [81]. This showed the feasibility of cryogenic antennae. Seismic, acoustic and thermal disturbances were reasonably under control. The group felt ready to take the next step towards developing a large antenna again.

Pizzella decided to install and operate it at CERN. An agreement was signed in which CERN would contribute with the buildings, cryogenic liquids and technical support. At the beginning of October 1980 the Explorer cryostat started to be delivered at CERN [73].

The Australian cryogenic Niobe project began in 1976 when David Blair came to University of Western Australia (UWA). The UWA project was an informal collaboration with Stanford and LSU, initially led by Prof Roy Rand, from SLAC, who came to UWA for 2 years. LSU supplied a small 6 kg niobium bar which was made into a magnetically levitated system that achieved a noise temperature of 40 mk, a record in those days. The Australian group, then, developed a 1 m levitated niobium bar, which operated inside the Niobe dewar. They then went on to get the worlds biggest piece of niobium around 1982 and it took 10 years to get cryostat, bar and suspension into working order. They had worked out that there was a maximum diameter that could be levitated (about 200mm) and they abandoned magnetic levitation in favor of a multimode suspension. Around 1987 they demonstrated a Q factor of 240 million, a world record for any metal [82, 83].

The Tokyo University group, which was already a well established GW detection group operating room temperature detectors, constructed two cryogenic resonant-mass detectors for the search of low frequency continuous GW wave signals coming from the Crab pulsar, in collaboration with KEK (High Energy Accelerator Research Organization). They were:
- CRAB III: 74kg, 5056 aluminum alloy, 60Hz torsion type antenna at 4.2K, which reached a sensitivity of h ~ 2 x $10^{-21}$ for 1,800 hours of observation [84];
- CRAB IV: 1200kg, 5056 aluminum alloy, 60Hz torsion type antenna at 4.2K, which reached a sensitivity of h ~ 2 x $10^{-22}$ for 1,900 hours of observation [85, 86].

The group at Moscow State University was working with cryogenic very high Q (~2 x $10^9$) silicon antennae with mass around 10 kg, and equipped with parametric transducers operating in the GHz region [48]. There was another one in Rochester (USA), which was using a 1.2 m long and 0.3 m in diameter 5056 Al cylinder of 208 kg, resonant at 2302 Hz, and cooled to 1.6-4.2 K [87], but they had discontinued operation by the mid 80's. The University of Maryland also built a cryogenic bar (M = 1400 Kg, L = 1.5 m, 6061 Al, f ~1700Hz) [48].

There was also a group in Regina (Canada). The antenna consisted of a single crystal of quartz of dimensions 30.5 cm by 2.5 cm by 1.9 cm of rectangular cross section, 385g of mass, resonant at 8846 Hz, with a mechanical Q ~ 3.3 x $10^7$ at T = 1K. The crystal was gold plated at two opposite faces and soldered to the plates in the nodal plane were four headed wires from which the crystal was suspended and through which electrical contact was made. Because of the piezoelectric properties of quartz, the antenna was its own transducer. The crystal was cooled to a final temperature of 3 x $10^{-3}$ K in stages: liquid N, liquid He and three stages of adiabatic demagnetization. The detector was operated in the period 1979-82 with a sensitivity of 2 x $10^7$ erg $cm^{-2}$ $Hz^{-1}$ at a temperature T ~ 1K, at which temperature the dominant noise was the noise temperature of the then available RF SQUID amplifier. Attempts to lower the RF SQUID noise temperature by constructing and adding a DC SQUID in front of the RF SQUID were not successful and the expected ultimate quantum limited sensitivity of ~ $10^4$ $cm^{-2}$ $Hz^{-1}$ could not be reached. The experiment was abandoned shortly after [88, 89].

The first bar antenna to reach a noise temperature below 50 mK was the one in Stanford in 1981 [90]. Perhaps one can say it was the first cryogenic resonant-mass antenna which was intended to be a gravitational wave detector to enter into operation. This antenna recorded data with an average noise temperature of 20 mK for about 74 days. However this performance was never repeated in the following years. In 1986, for example, the noise temperature was around 50 mK.

Explorer began operation in November 1985 and the LSU antenna (later called ALEGRO) started operation in 1986 [91, 92].

Finally, after one and a half decades of development, involving resonant superconducting transducers, RF SQUIDs, innovative vibration isolation systems, the first gravitational wave coincidence experiment between cryogenic resonant-mass detectors was finally achieved in 1986 [93]. The groups involved were Louisiana, Rome, and Stanford. The bars were Explorer (at CERN, Geneva), the later called Allegro (in Baton Rouge), and the 4800 kg Stanford bar (in Palo Alto). The data analyzed were collected during the period of April to July 1986. Unfortunately, the three antennae were still far from the design sensitivity, they were also too vulnerable to non thermal noise, and the total time of triple simultaneous operation was only 55 hours (60 hours according to my calculations [94]). The Chinese group, with a room temperature detector in Guangzhou, also sent data to LSU to be analyzed together with the data from the three cryogenic detectors. No triple coincidences were found at all thresholds for the data exchanged and no zero time delay bins were above three standard deviations, for any combination of two-detector data (among all four detectors). If real events were recorded in that period, they did not happened during these ~60 hours (by the way, Guangzhou was not operating during this period) and they were not numerous enough to show up in the combinations of any pair of detector data. In spite of the null result, this coincidence experiment was a milestone in gravitational wave research.

Paradoxically the successful 1986 coincidence experiment between the three cryogenic antennae was the reason why they were unprepared for the Supernova 1987A event in February of the following year. Well aware of the improvements the three groups should carry out on their detectors in order to increase their sensitivity, to diminish non-thermal noises (non-stationary or non-Gaussian noises), and to increase duty cycle (time operational/total time), they all decided to stop operation and go for the improvements. During the year of 1987, the LSU group, for example, completely redesigned the antenna support system, both in the low temperature region and on the outside of the dewar, in an attempt to minimize up-conversion (low frequency motion producing high frequency mechanical noise by non linear processes, such as friction) and to eliminate internal resonances near the antenna frequencies. They also stiffened the vibration isolation stacks, redesigned the shape and dimensions of their composed masses. The previous loose support allowed large amplitude low frequency motions to be excited, which were believed to be the cause of non stationary noise in the data sets. A new antenna suspension system was introduced, now using a single 7 mm diameter titanium vanadium rod about the center of mass, which was attached at two points to a 2.4 ton bronze intermediate H-shape mass to act as a stable mechanical ground for the antenna. They also added an extra liquid helium reservoir of capacity slightly under 600 liters, which increased the time required between helium transfers, and they introduced an aluminum mushroom shape calibrator transducer. Finally, they also changed the antenna itself to the high Q 5056 Al alloy [95].

The 50 kpc away supernova SN1987A event only found two room temperature bar antennae in operation: GEOGRAV (Rome) and Maryland. Even though the expected sensitivity was not high for gravitational waves, a correlation was claimed between the two room temperature bars and the three neutrino detectors: Mont Blanc, Kamiokande and IMB around 2:40 U.T., five hours before the strong neutrino coincidence between Kamiokande and IMB (the famous 8+11 neutrinos) [96, 77, 97, 98, 99]. Even if the correlation had something to do with SN1987A, many questions have yet no answer. Which neutrinos were detected around 2:40 U.T.? Were they part of a pre-phase explosion? Did the bars detected gravitons from SN1987A or some other particle that excited the bars by thermoelastic processes? In any case, we hope another supernova will solve this problem.

Contrary to 1986, which was a great year for the second generation resonant-mass detectors, 1989 was a bad year: Amaldi and Fairbank died, and the Loma Prieto earthquake badly damaged the Stanford detector. These events were decisive for the future of the Stanford group. In the beginning they were unable to fix the 4.2 K detector, because the group was completely busy, involved in the construction of an ultracryogenic antenna at that time. Later, around 1994 or 1995, it became clear to NSF (the US National Science Foundation) that it would not make sense to support the construction of any resonant-mass detector in an earthquake area. This was a dead sentence to the Stanford ultra-low temperature antenna (figure 4.3). Furthermore, it was complicate to start a new ultracryogenic resonant-mass project (bar or sphere). A new world order was under way in the 90's (cold war has ended, among other things). This and other boundary conditions were pressing NSF in the direction of Big Science. The opening of other ultracryogenic resonant-mass projects, such as the LSU TIGA (Truncated Icosahedron Gravitational Antenna) would have to obtain, after a very well elaborated management plan, necessary agreements for operation of multiple facilities as single experiment and to identify a strong and experienced project manager, as Richard Isaacson made clear in a Gravity Co-op meeting in Baton Rouge in April 7$^{th}$, 1995 [100]. Furthermore, it would be very difficult for NSF at that time to approve two new big gravitational wave detection projects, such as LIGO and TIGA. The US physics and astronomy community would not accept that. So, only LIGO was approved, and new projects for ultracryogenic antennae would have to be done as isolate initiatives in other countries, even though in close collaborations with each other.

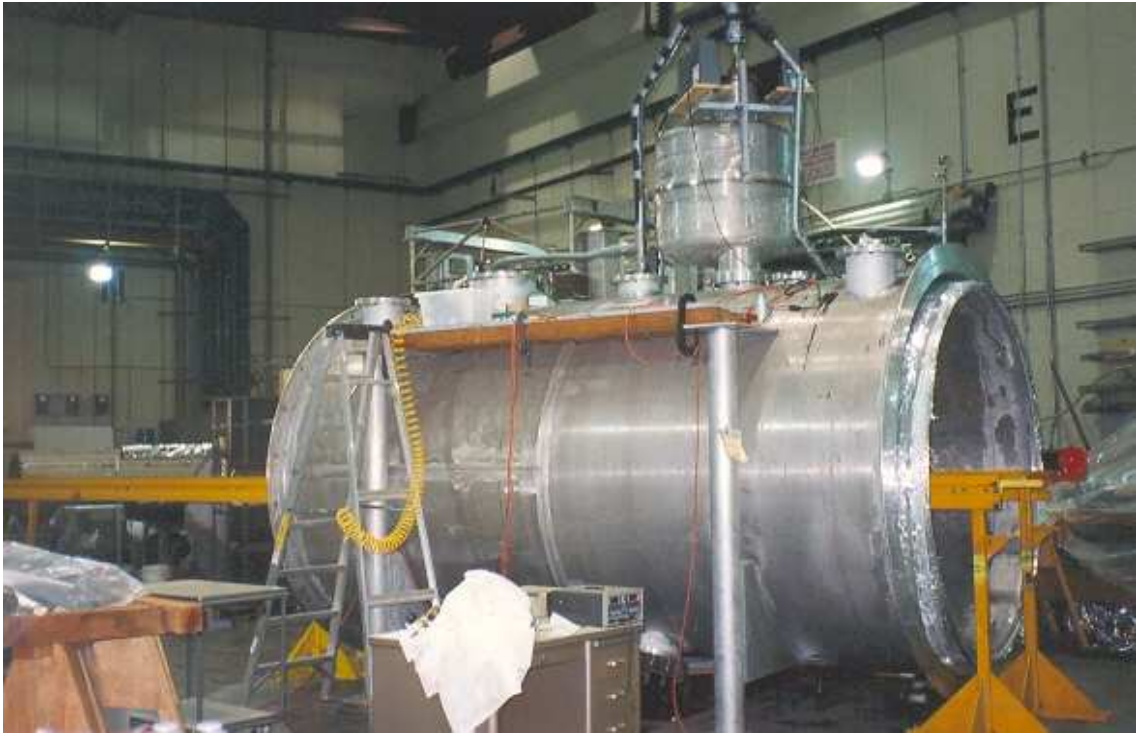

**Figure 4.3.** The Stanford ultra-low temperature detector which was proposed in the late 60's by Fairbank, but never completed (picture taken during the Marcell Grossmann Meeting at Stanford, Palo Alto, in 1994) (author's private collection).

## 5. The third resonant-mass generation and the regular operation of the bar network in the 90's,

Italy constructed the only two ultracryogenic bar antennae that ever went into operation: Nautilus and Auriga, showed in figures 5.1 and 5.2, respectively. The construction was part of the original 1969 Fairbank plan to cool the bar at 3 mK (an initially American plan that the American groups could not ultimately accomplish). These two bars were "twins", in the sense they were constructed from the same antenna and cryostat (dewar) design. Nautilus, named in honor of Jules Verne's submarine ship, had its external shell painted green, and Auriga was painted yellow. Nautilus came first. Its initial tests at 4.2 kelvin were performed at CERN during 1989, and in February 1991, the first ultra-low temperature ($< 0.1$ K) test of Nautilus of the Rome group was performed in the CERN laboratory [101]. In 1992 Nautilus was moved to Frascati [102] and has stayed there ever since. Then, in July 1994, it was cooled again to ultra-low temperatures. On the other hand, Auriga's lab was completed in 1992. The first cryogenic runs (using FET amplifier instead SQUID) happened between mid 1995 to mid 1996, and the first ultra-cryogenic run began in February 1997 (scientific data started in May 1997) [103].

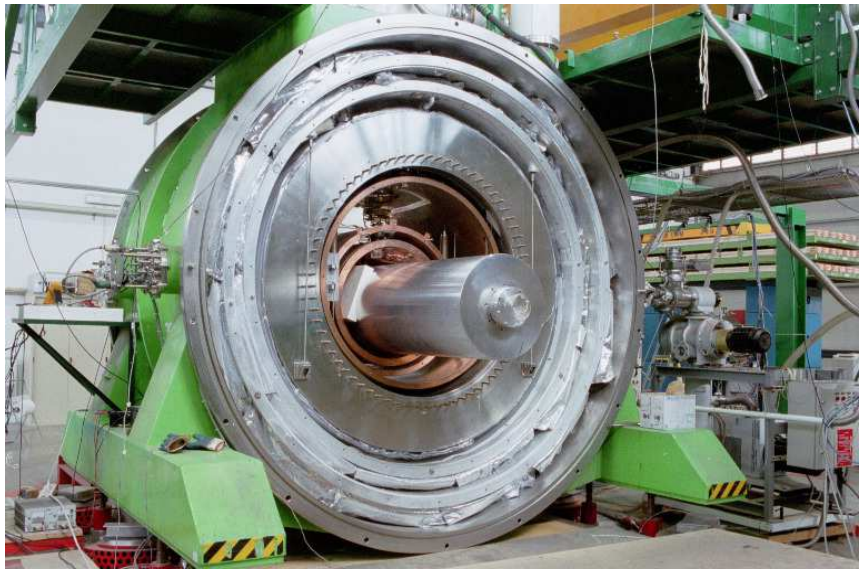

**Figure 5.1.** Picture of the Nautilus detector (open) (courtesy of the ROG collaboration).

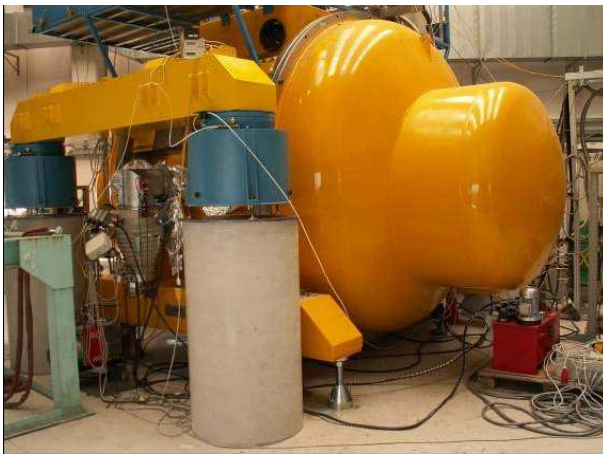 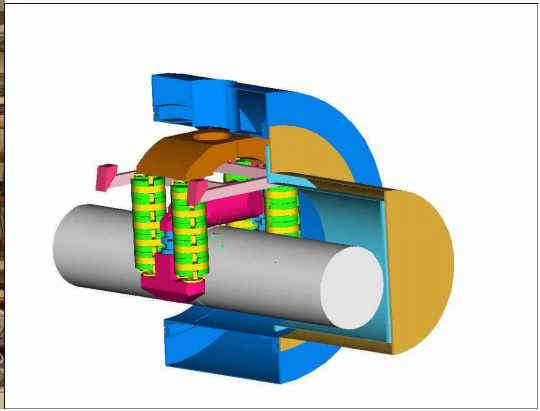

(a)  (b)

**Figure 5.2.** a) Picture of the Auriga cryostat equipped with the room temperature commercial gas damper (blue colour); b) Schematic of the Auriga detector partially sectioned. The toroidal liquid helium vessel is sky blue, the bar in gray and the column assembly of the cryogenic suspension is drawn in green and yellow (courtesy of the Auriga collaboration).

The 90's were the golden years for resonant-mass bar detectors already constructed. Most of the large baseline interferometers were under construction with one exception: TAMA300, which became operational in the beginning of 1999 [104], so most of the gravitational wave observations relied on the bar network of five state-of-art bars: Explorer (1990), Allegro (1991), Niobe (1993), Nautilus (December 1995), and Auriga (1997) (the year inside the parenthesis is when the detector became operational for scientific runs) [66, 105].

Even though the situation was good for the operation of bars it was not good for the proposal of new resonant-mass projects, in particular for sphere projects. There were already four project proposals (TIGA, GRAVITON, GRAIL, and SFERA) in elaboration for the construction of truncated icosahedron and/or sphere antennae when the "Cryogenic Gravitational Wave Antennae: a Progress Workshop", organized by

Massimo Cerdonio, took place at the INFN in Legnaro (Padova) in June 1993. However, none of them really became a reality before 2000 [106].

The major parameters for the five bar detectors mentioned are presented below [107, 108, 109].
 EXPLORER: 2270 kg Al antenna, located at CERN (Geneva, Switzerland), cooled at 2.6K, and equipped with a capacitive transducer and SQUID amplifier;
ALLEGRO: 2296 kg Al antenna, located in Baton Rouge (Louisiana, USA), cooled at 4.2K, and equipped with an inductive transducer and SQUID amplifier;
NIOBE: 1500 kg Nb antenna, located in Perth (Western Australia, Australia), cooled at 5K, and equipped with a parametric transducer and FET amplifier;
NAUTILUS: 2260 kg Al antenna, located in Frascati (Rome, Italy), cooled at 130mK with liquid helium dilution refrigerator, and equipped with a capacitive transducer and SQUID amplifier;
AURIGA: 2230 kg Al antenna, located in Legnaro (Padova, Italy), cooled at 200mK, and equipped with a capacitive transducer and SQUID amplifier.

Figure 5.3 shows the locations of the five bars, and the operational times of the network of detectors, from January 1997 up to June 2003 can be seen on figure 5.4. It is apparent that in that period of six and a half years (78 months) there were about two years for triple operation (three detectors on) and about four months for quadruple operation. Unfortunately, there was no period for the operation of all five bar detectors. However this figure demonstrated that bar detectors had achieved a reliable, high duty cycle and stable operation.

Today the network [107] has only three detectors (Explorer, Nautilus, and Auriga). Niobe and Allegro are no longer in operation.

Niobe was a bar, which had good sensitivity from the beginning. In 1993 the Australian group achieved about 4mk noise temperature and got the full parametric transducer system going reliably. Finally, on the last run in 2001, the group achieved about 600 microkelvin, but ran out of money and stopped operation. In that period they determined the low interaction of cosmic rays with the bar [82].

Allegro was a bar antenna with high sensitivity and duty cicle. It ceased operations in April 14, 2007, mainly because LIGO had surpassed its sensitivity around 900 Hz. Howerver, LIGO will be upgraded to the advanced version in the period of 2011-2013, so, perhaps Allegro should be back into operations during this three-year period.

The Nautilus and Explorer detectors with streamer tubes placed above and below the cryostats for cosmic ray detection/veto are shown in figure 5.5. The passage of cosmic rays has been observed to excite mechanical vibrations in resonant gravitational wave detectors with high sensibility [110], so these cosmic rays detections should be used for vetoing the times cosmic rays were registered.

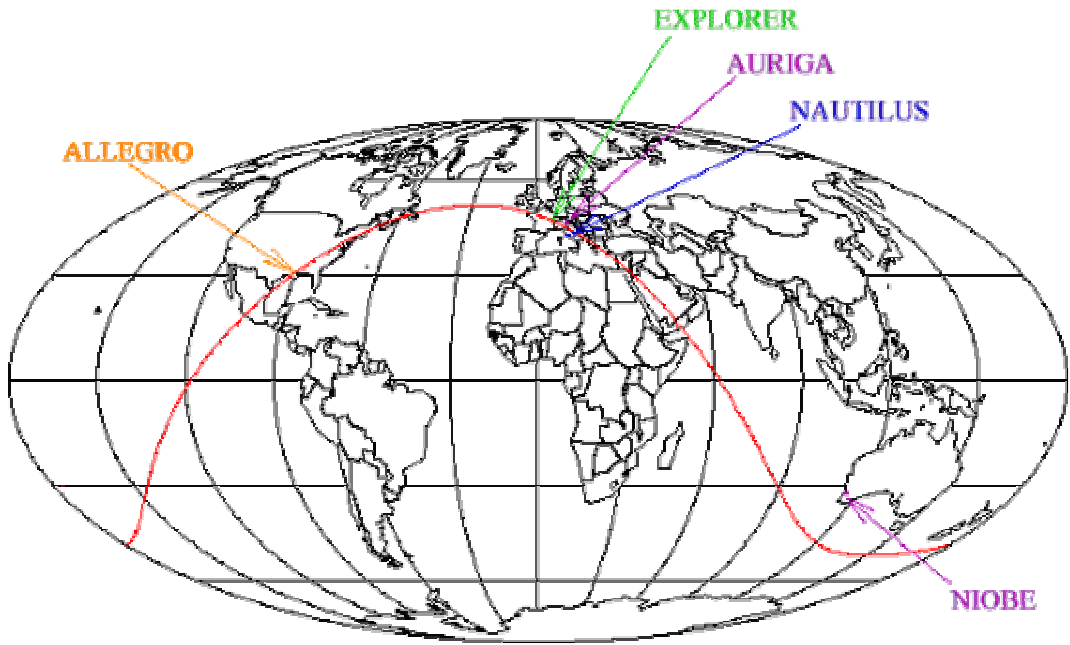

**Figure 5.3.** Locations of the five cryogenic bar detectors in the world which participate in the IGEC (International Gravitational Event Collaboration). All bars were usually kept oriented to the same direction in the sky. The way to do that was to keep them perpendicular to the Earth's great circle line shown in the map. This would optimize the GW signal correlation in energy, but would make them to loose the capability to determine the source direction (courtesy of the IGEC collaboration [107]).

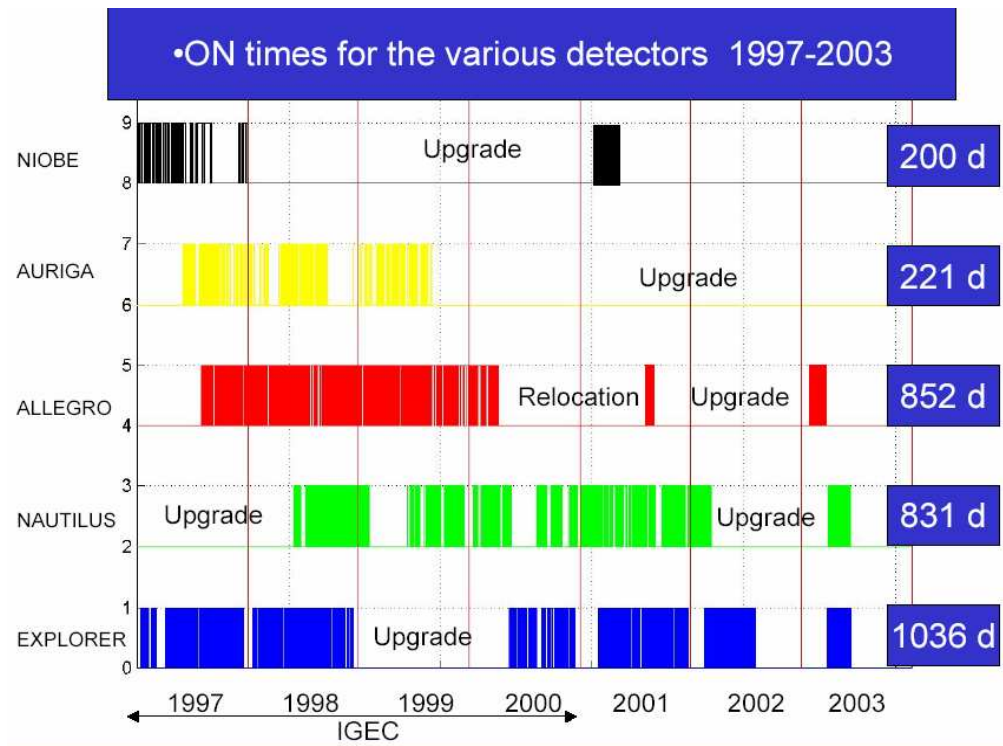

**Figure 5.4.** Operational times of the network of detectors, from January 1997 up to June 2003 (courtesy of the IGEC collaboration [107]).

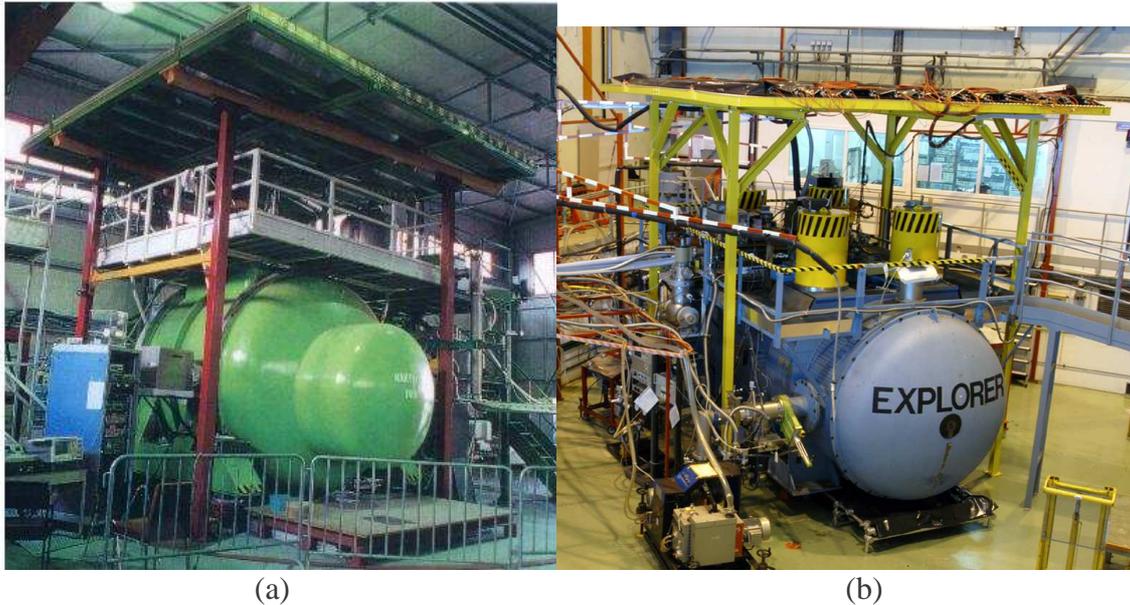

(a)          (b)

**Figure 5.5.** Nautilus (a) and Explorer (b) detectors with streamer tubes placed above and below the cryostats for cosmic ray detection/veto (courtesy of the ROG collaboration).

## 6. The present status of the resonant-mass gravitational wave detectors

It was clear during the time the interferometers were under construction or in commissioning phase that the bar antennae should (if enough money was available) be kept in operation. After all, there was no other way to try to detect GWs. However, we are now at a situation where a network of highly sensitivity interferometers is operational in scientific run mode. What will happen?

In order to answer this question we should ask ourselves the following question: what have the bar detectors achieved? One of the major achievements made by the bar detectors was about duty cycle and stable operation. Allegro has operated from February 2004 to 2006 with about 96% of duty cycle. Auriga has been presenting the same duty cycle (96%) since May 2005 (it actually resumed operation in December 2003). The duty cycles of Explorer and Nautilus were 87% and 86% in 2005, respectively, but in spite of that, in the period 2003-2004-2005 Explorer and Nautilus had an overlapping operation of 50% of the time on average (148.7 days, 218.5 days, and 182.1 days, respectively). Finally, Auriga-Explorer-Nautilus had a triple overlapping operation in 2005 covering 130.7 days from May 20 to Nov 15 (180 days), or about 72% of the time during that period [111, 112]. It will take some time before a network of interferometers reach a similar performance.

What about sensitivity? Are they good enough for scientific runs? Figures 6.1, 6.2, and 6.3 present the sensitivity curves of the current four operational bar detectors. These noise curves are a composition of a few noise contributions, such as antenna and transducer thermal noises, amplifier noise, and back action noise; therefore the cooling of the antenna from 4.2 K to 100mK does not necessarily improve the sensitivity curve if thermal noise, which is proportional to $\sqrt{T/Q_{mec}}$, is not the dominant one in the bandwidth. The initial results of the 2nd International Gravitational Event Collaboration

(IGEC2) are shown in figures 6.4, 6.5, 6.6, and 6.7. A significant progress was achieved in bandwidth increase for all four detectors in the last few years, proving that resonant-mass detectors can be broadband detectors when better transducers and amplifiers are used. There was also a significant improvement in sensitivity from IGEC1 to IGEC2.

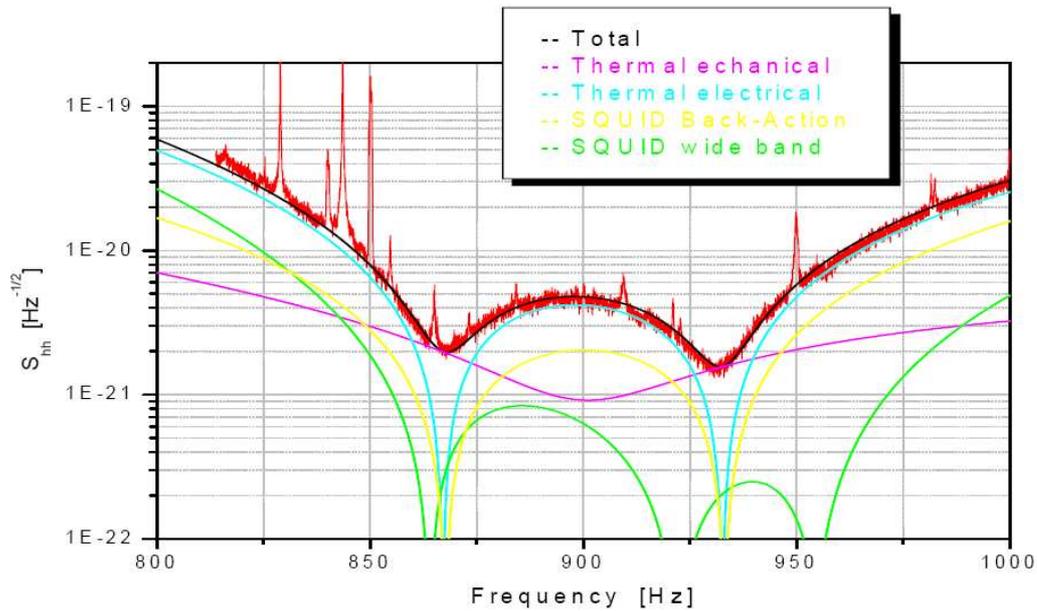

**Figure 6.1.** Power spectral density of the Auriga detector operating a 4.5 K, using a two stage low noise dc-SQUID amplifier. The agreement between the predicted value (black curve) and the experimental one (red curve) is quite good. The dominating noise source is the electrical resonator thermal noise (sky blue curve) except at the resonances, where the antenna thermal noise dominates (courtesy of the Auriga collaboration).

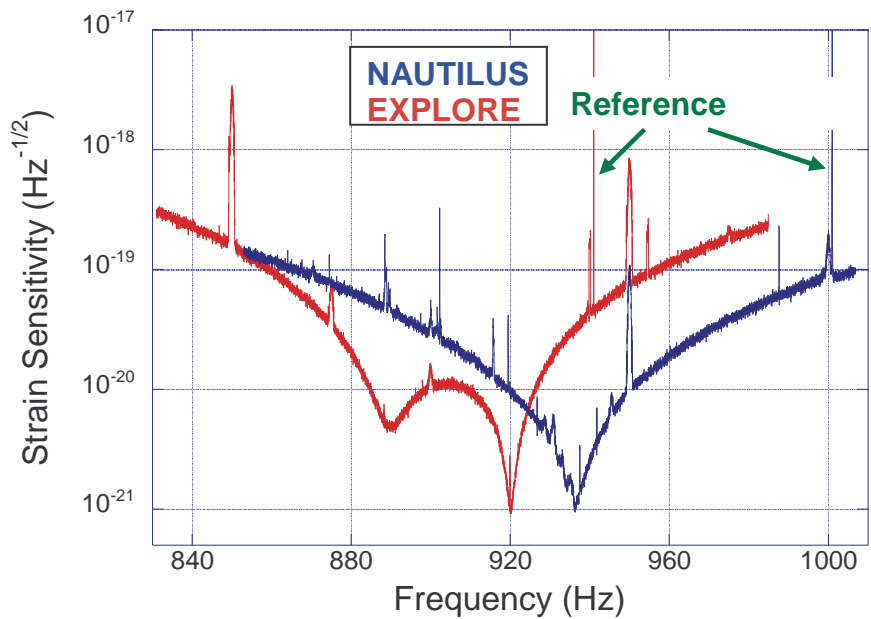

**Figure 6.2.** Power spectral density of Nautilus and Explorer detectors (courtesy of the ROG collaboration). The strain sensitivity is about $7 \times 10^{-22}$ Hz$^{-1/2}$ and the bandwidth is about 5 Hertz. The sensitivity for 1 ms bursts is $h = 3 \times 10^{-19}$. The Nautilus antenna was machined in order to reach maximum sensitivity around the SN1987A pulsar frequency.

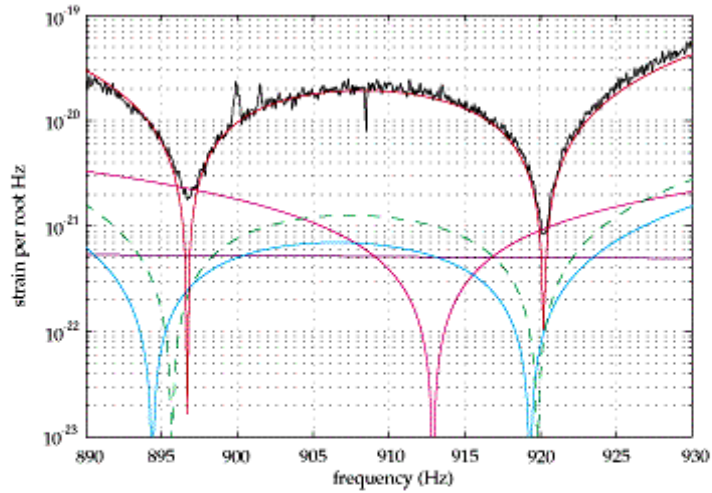

Measured strain noise spectral density of ALLEGRO and the various noise contributions which are predicted from the noise model of the detector.
— Measured total noise,  — antenna brownian,  — transducer brownian,
— transducer electrical loss,  — SQUID white noise,  — SQUID back action.

**Figure 6.3.** Measured strain noise spectral density of the Allegro detector using a two stage low noise dc-SQUID amplifier (courtesy of the LSU Gravity Allegro group). The SQUID white noise is the dominant one for the whole bandwidth except at the resonances, where the transducer Brownian dominates.

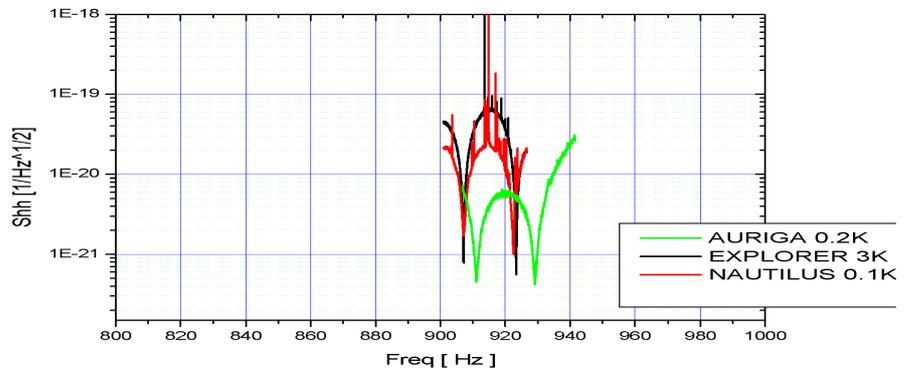

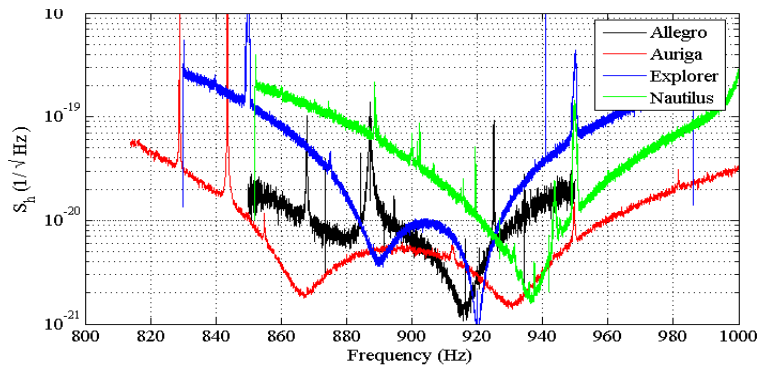

**Figure 6.4.** From IGEC1 to IGEC2 (courtesy of the IGEC2 collaboration). A significant progress was achieved in bandwidth increase for all four detectors in the last few years, proving that resonant-mass detectors can be broadband detectors with better transducers and amplifiers.

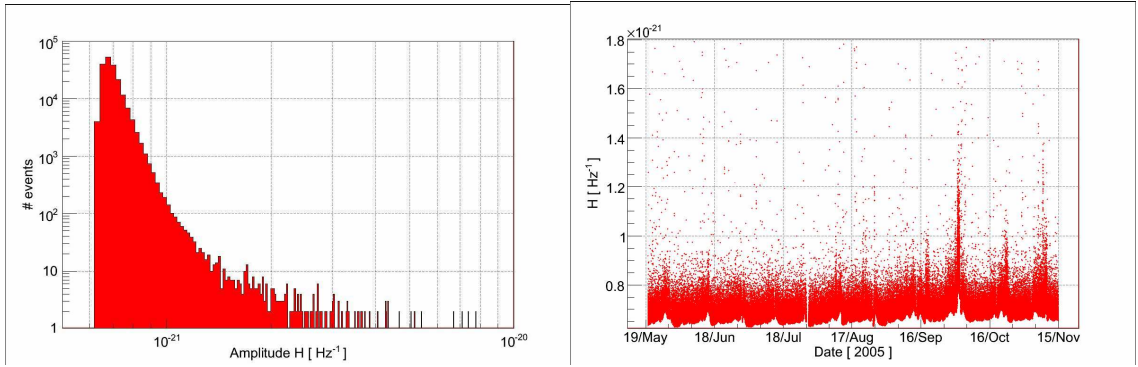

**Figure 6.5.** a) Cumulative histogram for impulsive events search for 173 days (May-November 2005) of Auriga data. The threshold is SNR = 4.5 and the total number of events 186911; b) Strip chart of the Auriga candidate events during the same period. It comes evident that the detector works in a very stable configuration with a small number of outlayers above SNR = 6 (12 par day). The short breaks occurring every two weeks come from the ordinary cryogenic operations (liquid helium refill) (courtesy of the Auriga collaboration). Explorer and Nautilus present similar curves, but displaced a factor of 2.5 to higher Hs (strain in $Hz^{-1}$ units) in figures 6.5a and 6.5b.

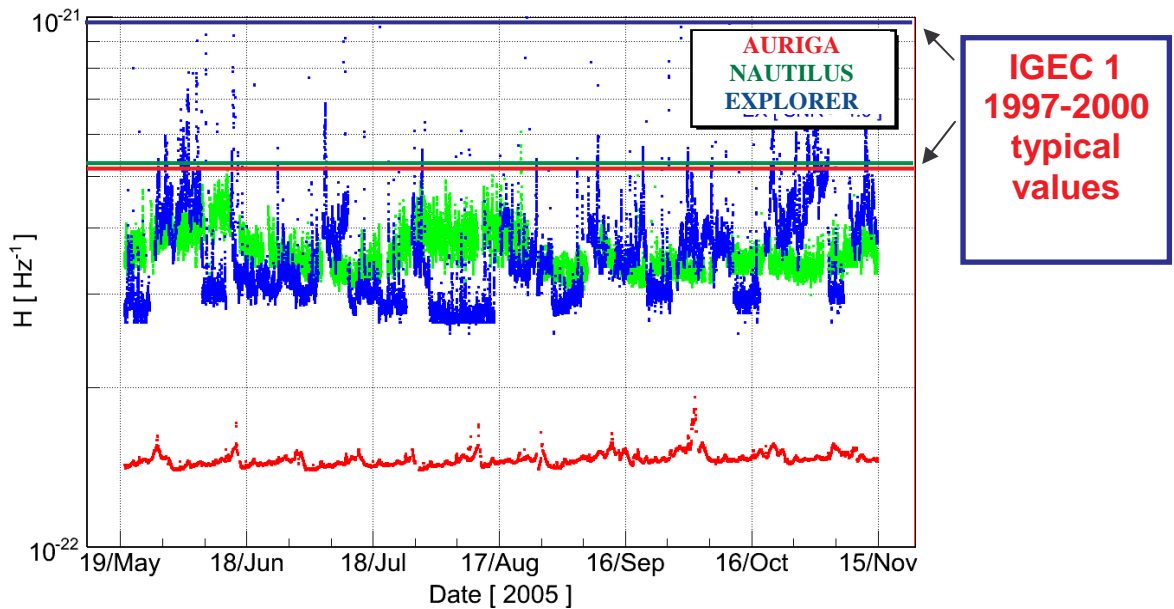

**Figure 6.6.** Noise of the Detectors is plotted versus Time in terms of the Fourier component H (courtesy of the IGEC2 collaboration). There was a significant improvement also in sensitivity from IGEC1 to IGEC2. The data exchanged between detectors considered thresholds around SNR > 4 for Explorer and Nautilus and above 4.5 for Auriga.

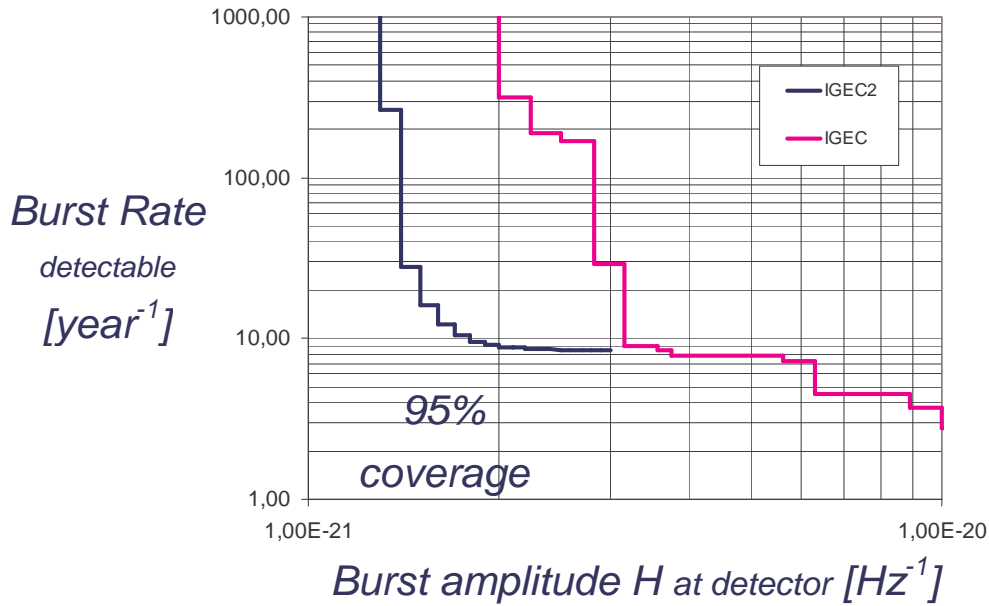

**Figure 6.7.** Burst rate as a function of burst amplitude (in $Hz^{-1}$) (courtesy of the IGEC2 collaboration). Only the regions on the left or under the curves are not eliminated.

IGEC2 was the only GW observatory in operation from May 20 to November 15, 2005 (180 days) involving all four bar antennae, which were co-aligned (pointing to the same region in the sky). It searched for a broader class of signals than IGEC1. Auriga, Explorer, and Nautilus exchanged data. The Allegro data was initially available for follow-up investigation only, but it will be added soon. The search for GW burst signals was tuned to identify single candidates with high confidence. Only triple coincidences, which had false alarm rate equal to 1/century, were searched. No candidates were found during the period. All four detectors presented high duty cycles and very low false alarms. Therefore, the identification of single candidates at low SNR was possible with very high confidence. The blind search (nobody knows the correct times of the event lists) in the statistical sense made the statistical interpretation non-controversial. IGEC2 also continued with four bars from November 15, 2005 until April 14, 2007, when Allegro ceased operation. Since then, IGEC2 continued with only the three Italian bars (Explorer, Nautilus, and Auriga). IGEC2 is still underway today (August 2009).

In spite of the outstanding performance of the current bar detectors, if one compares these sensitivity curves with the ones presented by the best laser interferometers performances around 900Hz, bars have been surpassed. In the overall, however, considering stability and duty cycle, they still should be kept operational, because a burst event related to GW detection can happen anytime. Therefore, continuous operation with high duty cycle is necessary. Furthermore, they would detect the wave from another physical principle: energy absorption from the wave, instead of direct geodetic deviation caused by its passage.

Can resonant-mass detectors do better in sensitivity? Can they reach the standard quantum limit?

The *standard quantum limit* (SQL) [70] is the minimum level of noise which can be obtained *without* the use of squeezed states. When the transducer is not able to

"squeeze" the signal (do the measurement with uncertainties in amplitude and phase with different magnitudes), the smallest value of measurement that one can do is the SQL. A resonant-mass antenna (bar or sphere) is a harmonic oscillation, whose amplitude and phase is sampled (measured) many times per second in order to detect the arrival of gravitational waves. If the transducer is of the passive type, such as used by all four current bars, in which a DC voltage or an DC superconducting current is running in the circuit, there is no way to control the *(quantum) demolition* one measurement produces in the immediately next measurement of amplitude and phase. If, however, an active transducer is used, such a parametric transducer, in which an AC power bias the circuit, it is possible to do *quantum non-demolition* (QND) [113] or, if the classical noises are still dominant, to do *back-action evasion* [114, 115]. Optical transducers [116, 117] or microwave parametric transducers [114] can circumvent this *standard quantum limit* and do a measurement with arbitrary precision. The "only" problem is to decrease the classical noises to the level of the *standard quantum limit* and to implement this artifice keeping the same stability and duty cycle bar detectors have achieved using passive transducers. So, better transducers and better amplifiers should be built if one wants to reach the *standard quantum limit* and surpasses it. This question is been addressed by the fourth generation of resonant-mass detectors: spheres and dual detectors, which will be presented in the next section.

## 7. The spherical antennae and other resonant-mass proposals for GW detection, and their prospects for the future

One can increase the sensitivity of a resonant-mass detector maximizing signal-to-noise ratio. Maximization of signal to noise ratio can go in two directions: the minimization of noise or the maximization of signal. If it is hard to decrease the noise, one can try to increase the signal. A GW antenna with a spherical shape (massive or hollow) seems to be the best solution for a given choice of frequency; because it maximizes the GW absorption (the transformation of gravitons into phonons) and is omnidirectional (it has equal sensitivity in any direction). Robert Forward, one of Weber's formers graduate students in the early 60's, was the first to realize this [69]. Almost nobody paid any attention to his idea at the time, as Weber had claimed in 1969 that coincidences had been observed. In 1971, all the groups involved in GW detection either had already started the construction of their bar antennae or had already submitted their proposals, in both cases they were willing to repeat Weber's experiment as close as possible, using bars. Only a few researchers (Neil Ashby, Joseph Dreitlein, Robert Wagoner, and Ho Jung Paik) [118, 119] performed some theoretical speculations about this new shape for a GW resonant-mass antenna in the years that followed. Then, ten years passed before Warren Johnson, who had already moved to LSU (from Rochester), revived Forward's idea in 1988. In 1989, LSU presented it at GR12, in Boulder, Colorado [120, 121]. It would have been nice if NSF had supported a spherical antenna project at that time (1989). I personally talked to Richard Isaacson about this at the conference cocktail, who acknowledged that a GW spherical antenna was an old idea (I will never forget that he was wearing Bermudas, black shoes, and black socks). Unfortunately this was another brilliant American idea along with Fairbank's ultra-low temperature bar, neither of which was ever implemented in USA.

In 1988 LSU started a solid line of research about spherical antennae, actually polyhedron antennae (in order to facilitate the transducer attachment on the antenna surface). Norbert Solomonson and I were too busy, involved with our transducers development, so Warren and Bill Hamilton had to rely on the work of undergraduate students in the beginning. When Stephen Merkowitz, who graduated from the University of Colorado and who had attended GR12, joined LSU as a graduate student things starting getting better. Warren and Stephen discovered the minimum transducer arrangement, which would make the antenna omnidirectional, and called it TIGA: Truncated Icosahedron Gravitational Antenna. The arrangement was to place six transducers on six of non-opposite pentagonal faces of a truncated icosahedron [122]. They demonstrated that TIGA could absorb 56 times more energy from a passing gravitational wave than the "equivalent" bar. Figure 7.1 shows the first room temperature TIGA constructed by LSU. Another twin TIGA went to the Rome University group. Both were machined from an old LSU 6063 aluminum bar. Their diameters were 84 cm [123].

Soon after finishing my Ph.D. program at LSU in December 1990 (I was officially on leave from the *Instituto Nacional de Pesquisas Espaciais* - Inpe (National Space Research Institute) since 1984 to develop such program), I began talking about starting a gravitational wave detection project involving the construction of a spherical antenna in Brazil. Laser interferometers were too expensive for the country, a fact I had already realized in the early 80's, when received many letters from experimentalists of both techniques (bars and lasers) mentioning detector costs for construction. I received the green light from Inpe around March 1991 to get involved with this research activity. One of the first people I talked to about this project was the editor of this book, Prof. José Antônio de Freitas Pacheco, who gave me a list of people involved with the Virgo laser interferometer collaboration in Brazil since October 1989. They were Mauro Cattani, Armando Turtelli, Nilton O. Santos, Carlos Escobar, and himself. He also gave me a longer list which includes people that I could talk about my idea, such as Jorge Horvath and José Carlos N. de Araujo. Pacheco and Escobar had at that time two doctorate students: Nadja S. Magalhães and Walter Velloso, who joined the Graviton group. Many others followed. The name Graviton, which is suitable for a detector whose principle of operation is to absorb quanta of energy from the GWs, came from a meeting at Mario Novello's house in Rio around February 1991. Luis Alberto R. de Oliveira suggested this name.

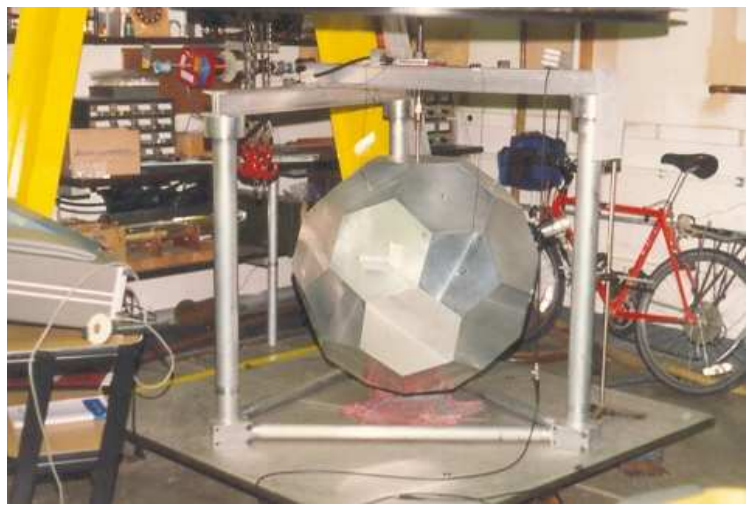

**Figure 7.1.** The LSU TIGA, with which Stephen Merkowitz did his thesis work (his bike gives an idea of the size of the antenna) (author's private collection).

Carlos Escobar, heavily involved in the Auger project nowadays, arranged a meeting between Giorgio Frossati and I at his office in the University of São Paulo (USP) in January 1993. Giorgio, an expert in dilution refrigerator development and a respectable researcher in low temperature physics, who had lived in Brazil from 1947 to 1970, had been enthusiastic about gravitational waves since the publication of Weber's first papers. The cooling of a giant spherical antenna weighting dozens of tons was a challenge he relished facing and this was an opportunity for him to pursue this research in the Kammerlingh Onnes Laboratory, in the Netherlands. Five months later he had already presented the proposal for the construction of Grail, a 30ton-2.6m in diameter aluminum alloy spherical antenna cooled to 10 mK [124, 125, 126, 127], at Legnaro workshop (organized by Massimo Cerdonio). This had a cascade effect. The contacts he made before the workshop with the Rome group motivated it to submit a proposal for Sfera, a similar antenna, in Italy, which, in turn, stimulated LSU to propose Tiga to NSF in the US. The Graviton project, which had already been informed to the gravity community at the Marcel Grossmann Meeting in Kyoto in 1991 (by a letter of intention) and had been presented as a contribute paper (poster) at GR13, in 1992 [128], had to wait a few years, in order to study the project viability (including a search for a site) and to form a group, before being submitted to the Ministry of Science and Technology (MCT) and FAPESP in Brazil in 1996 and 1998, respectively.

During the 90's a lot of studies regarding spherical and polyhedron shape antennae were performed world-wide [122, 125, 129, 130, 131, 132, 133, 134, 135, 136, 137, 138, 139, 140, 141, 142, 143]. A detector with a spherical antenna can determine the wave polarization and localize its astrophysical source in the sky for all gravitational wave signals detected in any given gravitational theory [69, 144]. Furthermore, it is never "blind" to any particular direction or polarization of the arriving wave [131]. Both these advantages of a spherical antenna are due to its omnidirectionality achieved by the use of six transducers placed in accordance with the truncated icosahedron configuration monitoring the quadrupole modes plus a transducer monitoring the monopole "breathing" mode [122, 145]. Another very important advantage of this kind of detector is precisely due to fact that it has many transducers monitoring many quadrupolar modes. It is like having many detectors operating together at the same time and site [133]. So, you can perform real time data analysis looking for correlations with the signals of the transducers, which is impossible to do using detectors located at different sites.

The Dutch group in collaboration with the Rome group also made significant progress in finding a suitable material for the antenna with high density (8 $g/cm^3$), high mechanical Q (~2 x $10^7$), and non-superconductivity at ultra-low temperatures (~ 15 mK). The group tested many small spheres at these temperatures [134, 146]. Some of them are shown in figure 7.2.

The Grail project was rejected in the Netherlands, so, the Dutch group, in collaboration with the Rome group and the Brazilian group, decided to build Mini-Grail. This became the first spherical antenna for gravitational waves, when it started operation at ultra-low temperature (~80 mK) using three transducers, in 2004. The second spherical antenna to go into operation was Schenberg at 5 K from the Brazilian group, in September 2006, also with three transducers. Both detectors are now in the commissioning phase and are expected to begin scientific run soon.

The Dutch Mini-Grail detector is shown in figure 7.3. It is composed of a cryogenic 68 cm diameter spherical gravitational wave antenna made of CuAl(6%) alloy with a mass of 1400 Kg, a resonance frequency of 2.9 kHz and a bandwidth around 230 Hz [147, 148].

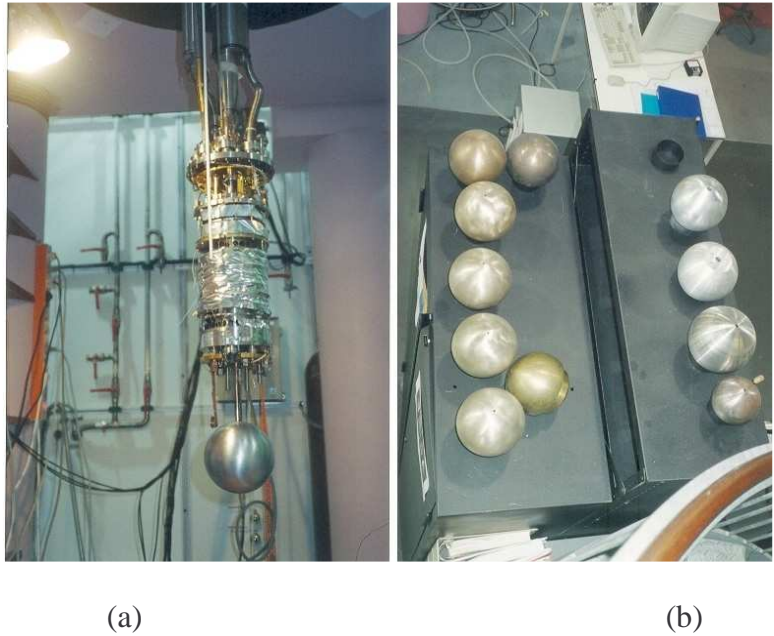

(a)                                        (b)

**Figure 7.2.** a) The Dutch setup (dilution refrigerator + sphere + hammer + PZT sensor) for measuring mechanical Q of small spheres (author's private collection); b) Some of the small spheres tested (author's private collection).

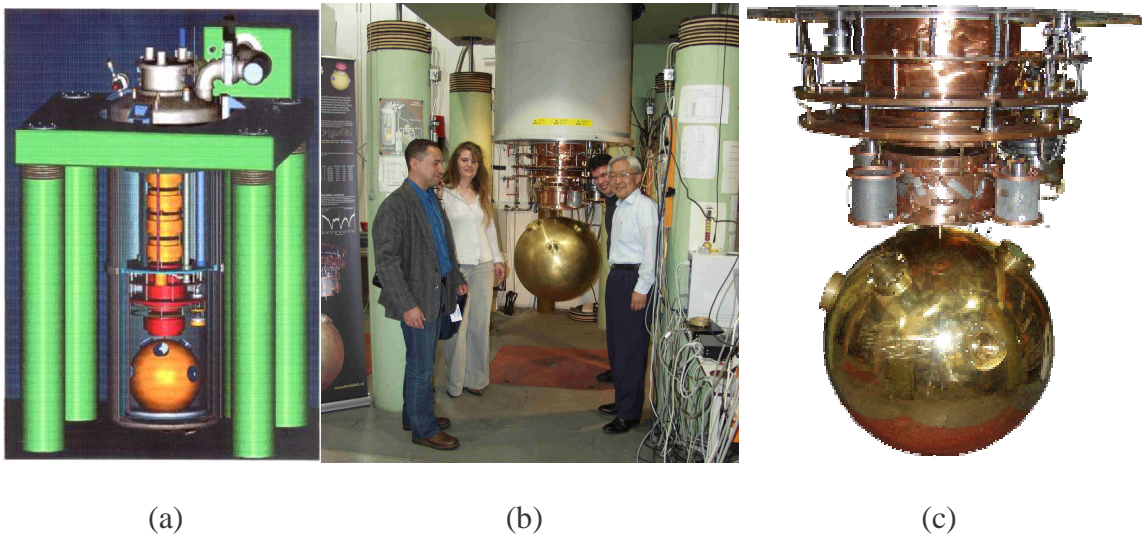

(a)                         (b)                         (c)

**Figure 7.3.** The Mini-Grail detector. a) Schematic view of the Mini-Grail detector. Six capacitive transducers distributed according to the truncated icosahedron (TIGA) arrangement proposed by Johnson and Merkowitz will monitor their fundamental modes of vibration; b) The detector lifted by the hydraulic system (from left to right: José Carlos N. de Araujo, Arlette de Waard, Odylio D. Aguiar, and Ho Jung Paik. Giorgio Frossati took the picture in July 2004; c) Detail of the Mini-Grail antenna equipped with a dilution refrigerator (courtesy of the Mini-Grail group [147]).

Figure 7.4 shows a schematic view of the Brazilian Mario Schenberg detector at the Physics Institute of the University of São Paulo, in São Paulo city [149]. The Schenberg CuAl6% antenna has a diameter of 65 cm and weighs 1.15 ton. The mechanical $Q$ (figure of merit) of the antenna is 2.7 million at $T = 2$ K, and it seems to follow the expression $Q = 4.9 \times 10^6 \, T^{-0.86}$ [21]. It has nine little holes on its surface for up to nine transducers, six of which follow the TIGA configuration[19]. At the *standard quantum limit* sensitivity it will have a strain noise power spectral density of ~$10^{-22}$ Hz$^{-1/2}$ [150]. Further improvement can be made because parametric transducers are able to squeeze signal and surpass the *standard quantum limit*. The construction was fully supported by FAPESP (the São Paulo State Foundation for Research Support) and started around 2001.

Both spherical detectors will operate in coincidence with each other, along with some long baseline laser interferometer detectors [151], searching for high frequency events in the 2.7-3.4 kHz frequency bandwidth. For the GW signals that contribute with power in both bandwidths (~900 Hz and ~3 kHz), such as some types of bursts, the coincidence is possible even with the bar detectors.

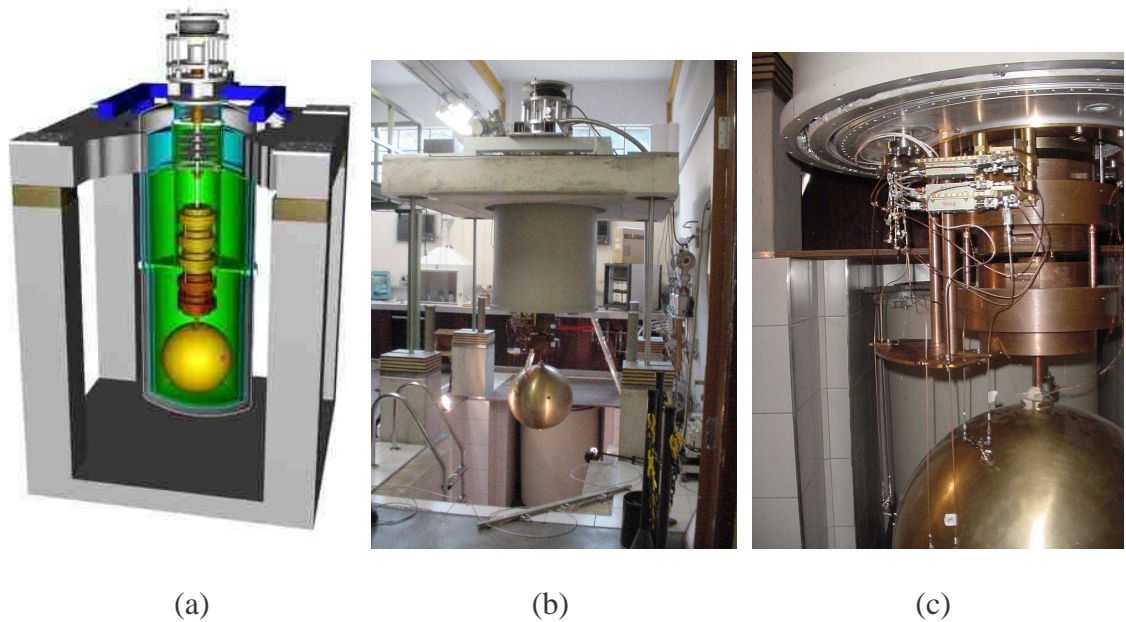

(a)     (b)     (c)

**Figure 7.4.** The Mario Schenberg detector: a) Schematic view of the Schenberg detector. Six parametric transducers distributed according to the truncated icosahedron arrangement proposed by Johnson and Merkowitz will monitor their fundamental modes of vibration; b) The Schenberg detector lifted by the hydraulic system; c) Detail of the antenna with three transducers installed. The three cryogenic HEMT amplifiers are on the top, just below the liquid helium reservoir.

Each one of the five quadrupole modes of the Schenberg sphere has an effective mass for quadrupole oscillation of 287 kg. These five independent modes will be coupled to the modes of six transducers, which have oscillating masses of 53.6 g and 10 mg. All these modes are tuned to the ~3.2 kHz resonant frequency and the energy flowing from the sphere modes to the 10 mg transducer masses due to this coupling produces an amplitude gain of about (287 kg / 10 mg)$^{1/2}$ ~ 5 k [68, 152]. The 10 mg transducer mass is the effective mass of a membrane that closes the microwave klystron cavity and forms a 0.1 mm gap with the top of the cavity post. This membrane is made of pure silicon (figure 7.5) and is supposed to have high mechanical Q.

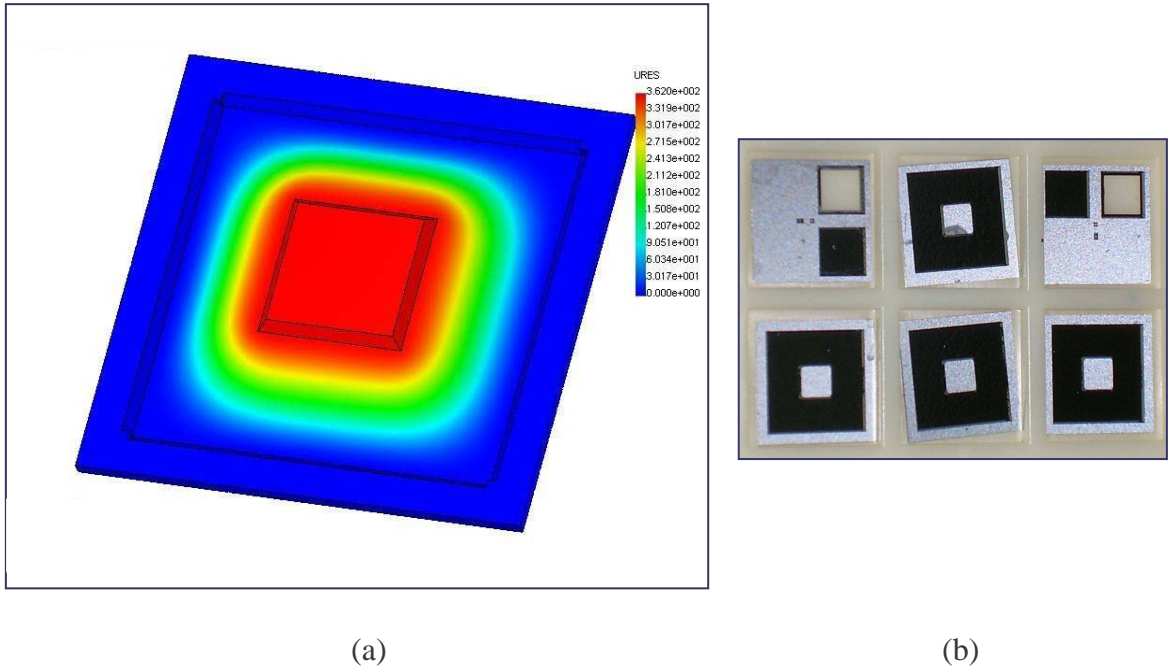

(a)                                                                              (b)

**Figure 7.5.** The silicon membrane: a) Photolithograph membrane design and results of a numeric simulation using the software Cosmos (the membrane oscillation is of its fundamental mode). The amplitude scale, in microns ($10^{-6}$ m), is shown on the right; b) Picture of the first membranes made. It is possible to produce them with a thickness of $20 \pm 1$ μm and with an island in the middle to adjust the value of the effective mass.

This kind of parametric transducer used by the Schenberg detector was studied and developed by both the Japanese [153, 154, 155] and the Australian [156] groups. In the same way as the Australians were doing with Niobe, the signal is sent to and received from the transducers by pairs of microstrip antennae, so, no wires or coaxial cables touch the spherical antenna (figure 7.6). The preamplifiers, which use HEMT technology, as the Australian group was planning to do with Niobe [157], have a noise temperature of 8-10 K, which corresponds to 12-14 $\hbar$ in sensitivity at 10 GHz ($T_n = 9$ K ~ 13 $\hbar \omega / k_B \ln 2$) [158]. However, as described above, we have changed the mechanical design of this transducer [159].

We believe this mechanical *innovation* in the transducer design was a good decision in order to increase the antenna sensitivity and to reach the *standard quantum limit*. [160, 113] The drastic change in the last mass from the Niobe bending flap mass, which was around 0.43 kg, to just 10 mg [161] gave this kind of transducer many advantages. Firstly, the pump requirement of phase noise to achieve the standard quantum limit of sensitivity dropped from −180 dBc.Hz$^{-1}$ to only −145 dBc.Hz$^{-1}$, which is a much more feasible goal. Secondly, a high electromechanical coupling can be accomplished with only a few nanowatts of pumping power, which is excellent news for one trying to cool the antenna down to the lowest thermodynamical temperature possible. Thirdly, the amplitude gain, which is related to the square root of the ratio between the last mass and the antenna effective mass, increases from ~26 to ~5.4 k. Finally, the need for carrier suppression for maximum HEMT performance (they require less than −80 dBm at the input [162]) will also decrease, because the power injected in the cavity is smaller.

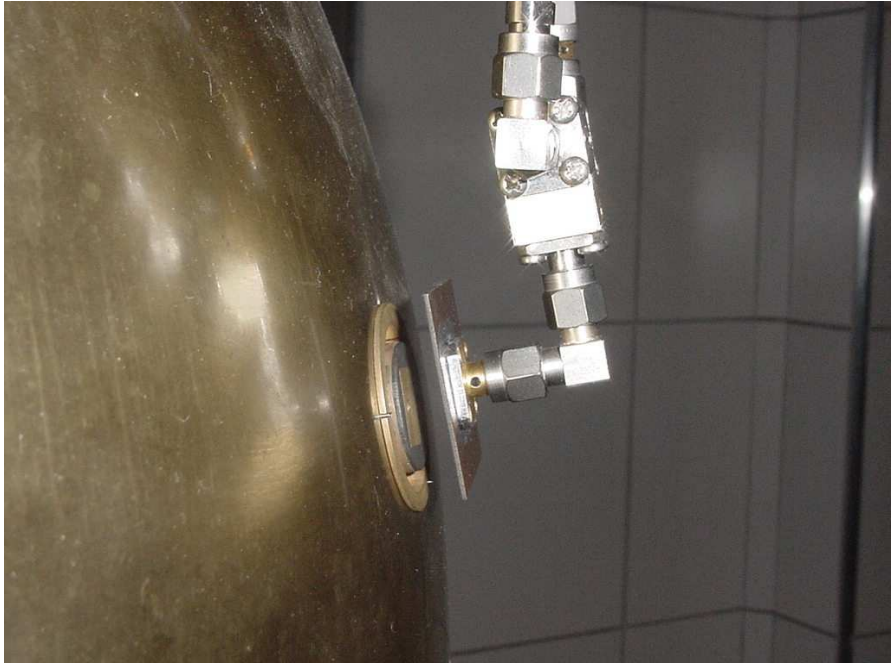

**Figure 7.6.** Detail of the Schenberg pair of microstrip antennae positioned in front of each other. The microwave pump signal from the very low phase noise oscillator jumps from the antenna on the right to the antenna on the left. The later sends the signal to the klystron cavity. Then, the modulated signal jumps back and, thanks to a circulator (which is just above the right angle plug-plug SMA connector), is sent to a cryogenic HEMT preamplifier.

Figure 7.7 shows the predicted strain noise spectral density considering conservative parameters and the Schenberg antenna operating at 4.2 K. The various components of the noise curve of one mode channel can be seen. Far away from the optimum sensitivity, the electronic noise dominates. At the optimum sensitivity, thermal noise dominates. The histogram of the optimal linear filter output for a short burst signal, using the best 12 hours of the first commissioning run of Schenberg, is shown in figure 7.8. It presented an absence of outliers outside the Gaussian noise distribution. However, its sensitivity is not as good as the outstanding sensitivity reached by Mini-Grail on its first commissioning run in 2004, which is shown in figure 7.9.

The map with the locations of the bars, interferometers, and spheres is presented in figure 7.10.

The expected sensitivity for a 2 meter sphere, such as the SFERA detector (SFERA: Proposal for a Spherical Gravitational Wave Detector, 2005), involving a collaboration between Italy, Switzerland and the Netherlands, is shown in figure 7.11. The curve represents the sensitivity that can be achieved with a read-out at the quantum limit. For a 20 $\hbar$ SQUID the sensitivity worsens by a factor of only 4-5. On the other hand, by the use of parametric transducers and squeezing of signal sensitivities better than $10^{-23}$ Hz$^{-1/2}$ can be achieved.

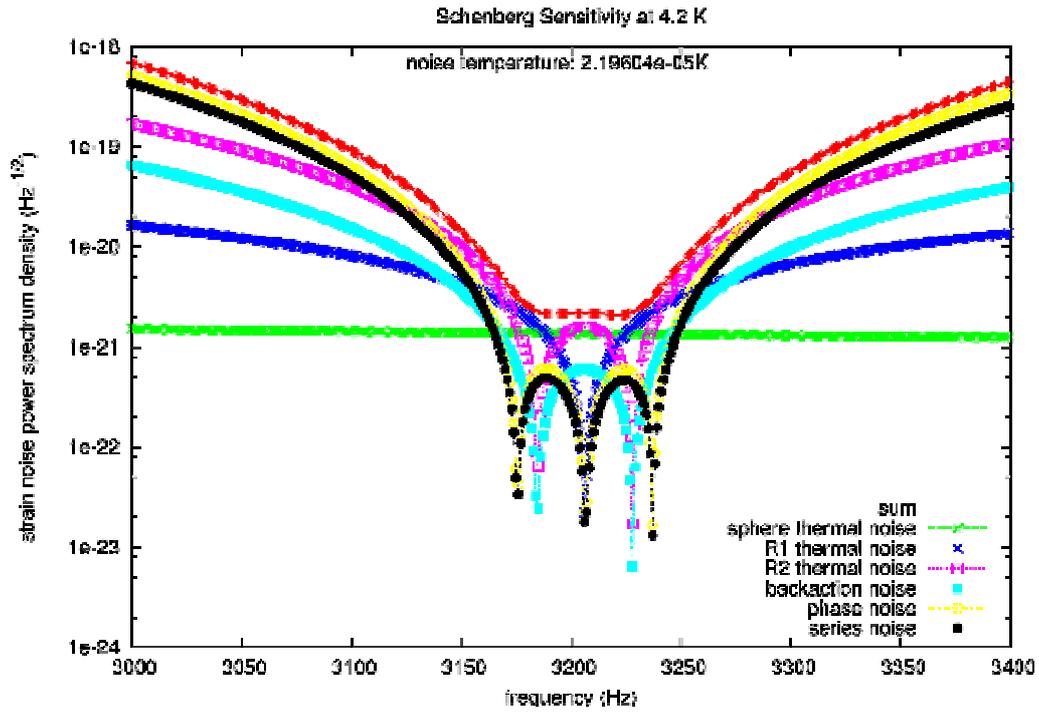

**Figure 7.7.** The Schenberg sensitivity curve at 4.2 K (red line) for a single quadrupole mode and the individual contributions of the noise sources.

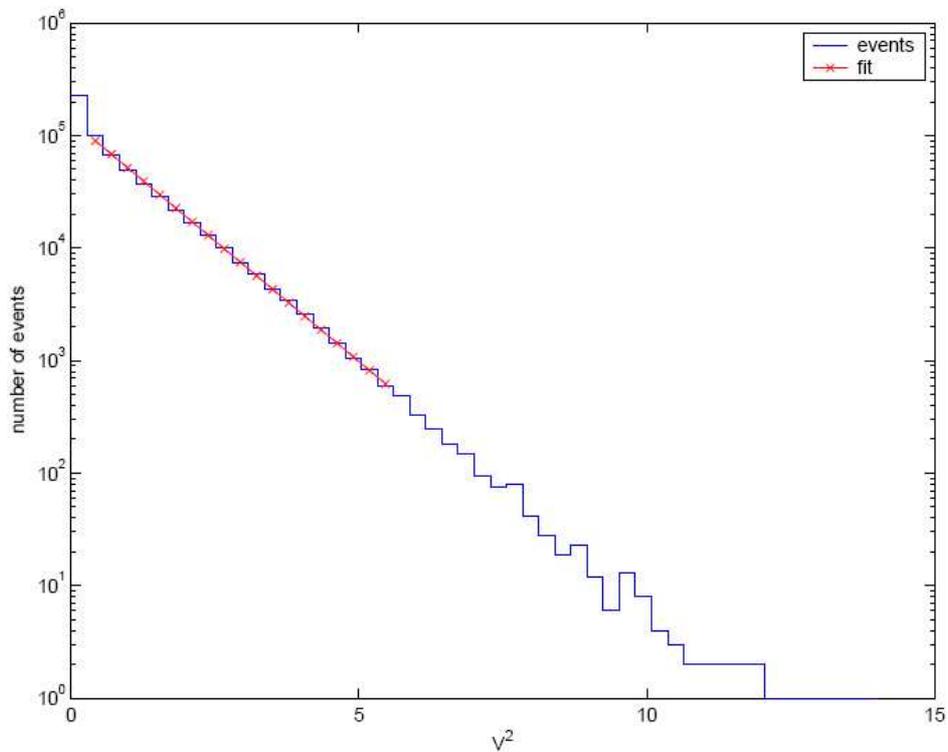

**Figure 7.8.** The histogram of the optimal linear filter output for a short burst signal using the best 12 hours of the first commissioning run of Schenberg. It presented an absence of outliers outside the Gaussian noise distribution.

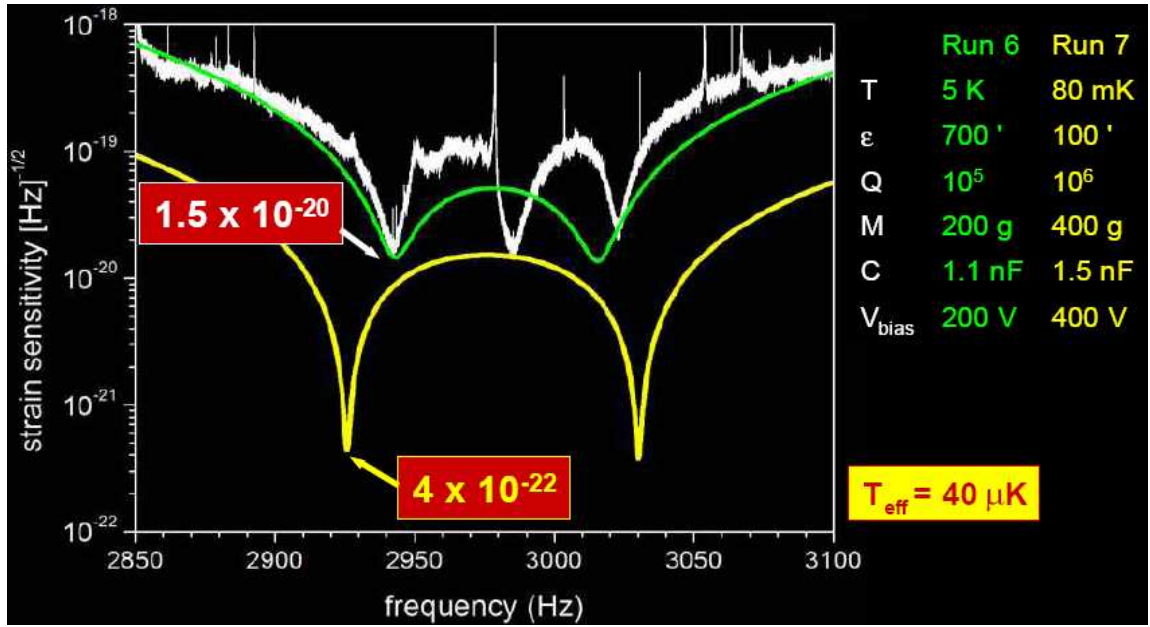

**Figure 7.9.** Mini-Grail sensitivity (white curve) in the first commissioning run (6th overall cryogenic run) and the prospects for the next 7th run (yellow curve) are shown (courtesy of the Mini-Grail group [163]).

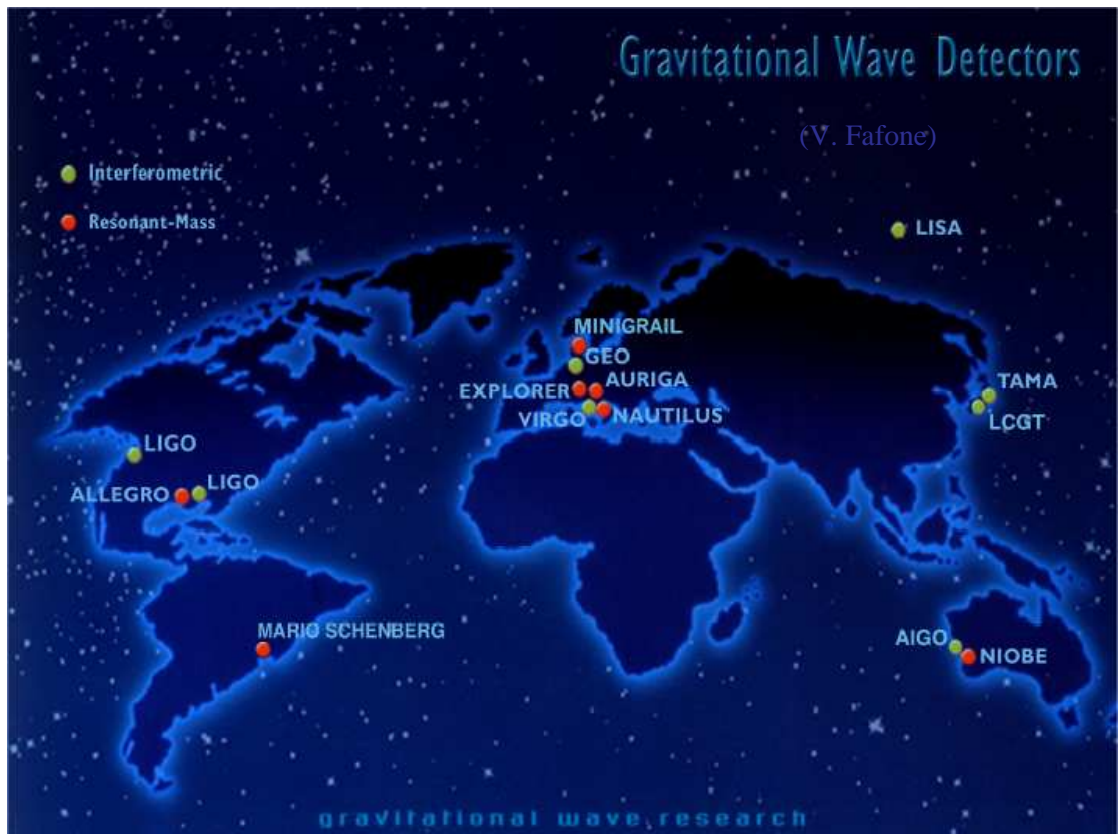

**Figure 7.10.** State of art cryogenic resonant-mass detectors and interferometers in the world (courtesy of the ROG collaboration).

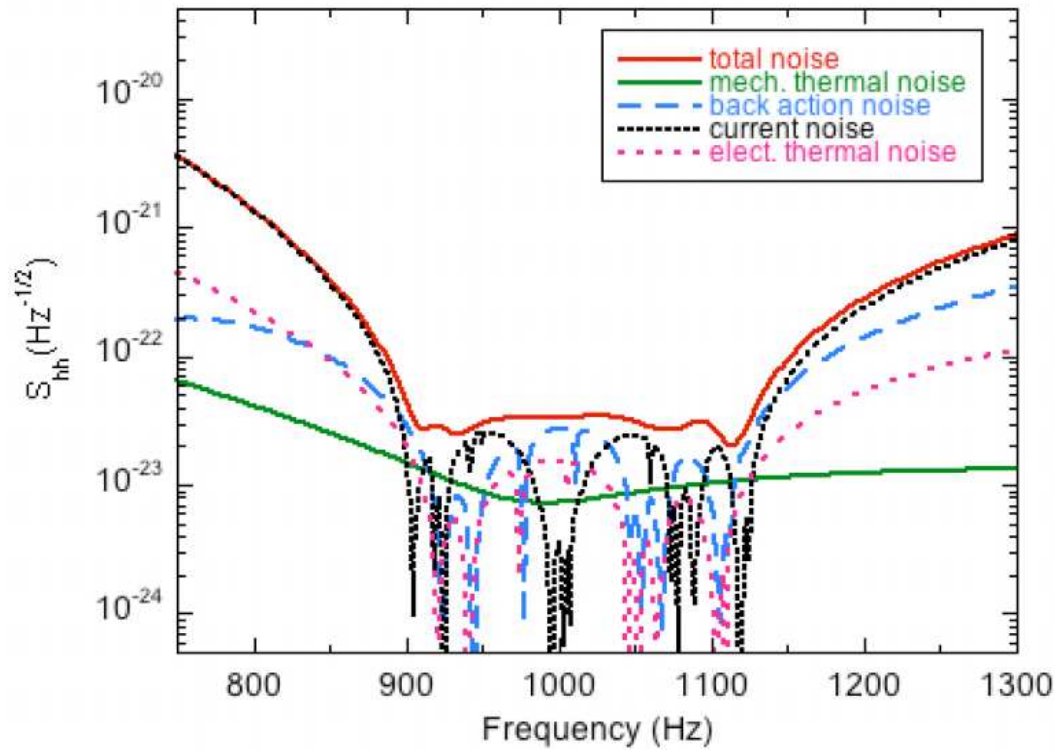

**Figure 7.11.** Calculated strain noise spectrum for a 2 meter diameter CuAl sphere. The sphere is equipped with six transduction chains in the TIGA configuration, with noise at quantum limit (SFERA proposal). For a 20 $\hbar$ SQUID the sensitivity worsens by a factor of only 4-5 (courtesy of the SFERA collaboration).

Spherical self-gravitating bodies, such as the Earth, the Earth's moon and other moons (Mars' e.g.) were also considered as gravitational wave antennae [118, 164, 165]. If the deployment of sensors on these bodies is technological and economical feasible it would be a way to search for GWs below 10 Hz. Interesting science could be done in this frequency bandwidth such as pulsar search, the study of binary systems of many types, super-massive black holes, and cosmological background search (coming from the inflationary period or from a pre-big bang universe) [13, 166].

The present resonant-mass detectors do not have as large bandwidths as the ones laser interferometers have. The multi-mode solution proposed by Richard in 1984 [68] or the more exotic Paik solution [167] have some practical limitations. An obvious way to circumvent this problem is by the use of a set of resonant-mass detectors each of them tuned to a different frequency in such a way to cover a large overall bandwidth. This is the so called "xylophone" solution [122, 137]. Another way to go is the Dual detector proposed by the Auriga group.

In a Dual detector the relative motion between two massive bodies must be measured, avoiding noise contributions from the lighter resonant transducer. A Dual detector can be formed by two nested massive bodies whose quadrupolar modes (i.e. the modes sensitive to the GW signal) resonate at different frequencies. In reference [168] the two resonators are two spheres, a full inner one and a hollow outer one. A dual cylinder configuration has also been evaluated [169]. These two configurations are shown in

Figure 7.12. The signal is read in the gap between the two masses as their differential deformations: the centers of mass of the two bodies coincide and remain mutually at rest while the masses resonate, thus providing for the rest frame of the measurement [170].

It can be shown that the response to GW excitation is preserved at any frequency between the resonances of the two test masses, while the response to the back action noise force, which acts out-of-phase on the two resonators, tends to be reduced (because it acts with opposite sign on the faced bodies). The sensitivity of the dual detector is predicted to be of interest in the frequency range between the first quadrupolar modes of the two masses. This can be as broad as a few kHz in the kHz range. Of course the transducer system is required to be wide band and suitably adapted, i.e. to provide the optimal balance between the displacement noise and the back-action force noise [170].

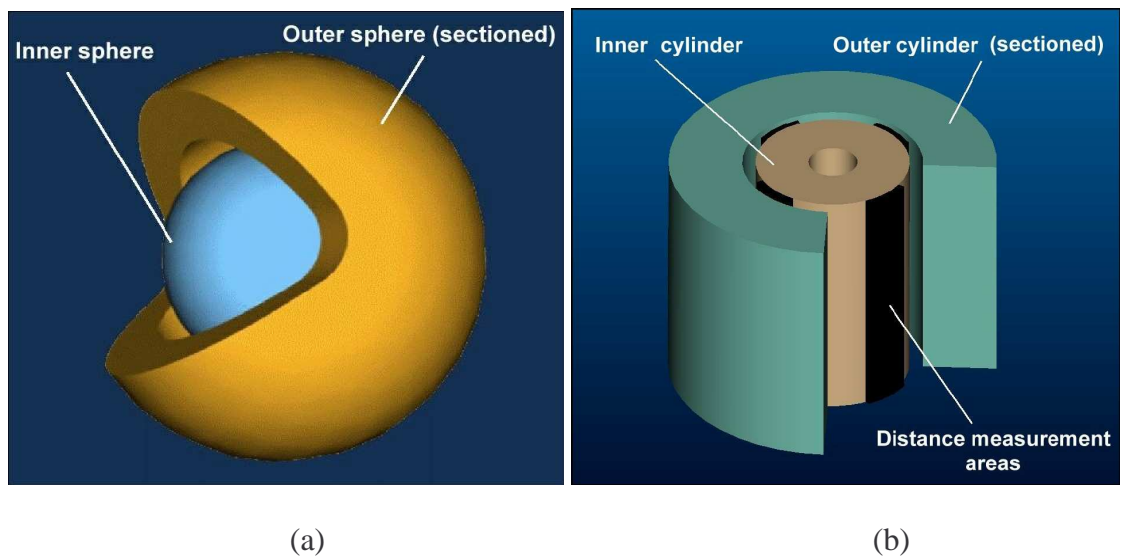

(a)                  (b)

**Figure 7.12.** Different Dual detector configurations are shown: a) the dual spheres; b) the dual cilinders (courtesy of the Auriga collaboration).

In figure 7.13 the sensitivity curves of two dual detectors can be seen. These curves could be considered as the maximal sensitivity possible with these detectors. The possibility to reach such results strongly depends on supportive researches, mainly on the material properties and on the readout sensitivity. Better sensitivities could be only obtained with larger detectors, which we consider hardly attainable with the present technologies.

The limited bandwidth of traditional acoustic detectors is due to the usage of the resonant transducer, which is needed to reduce the effect of the noise of the amplifier. In this case bandwidth enhancements can be obtained by the use of properly optimized readouts, but the thermal noise contribution of the resonant transducer limits the bandwidth to about 10% of the detector central frequency [170]. A revolutionary solution would be if one goes back to the old 70's idea of non-resonant transducers, but now using a "super transducer" with an ultra high sensitivity (ultra-high df/dx, for example) and very high electrical Q in order to decrease the influence of all the noises associated with the electrical and amplifier parts. In this case, the dominant noise would be the antenna Brownian, which is white and, so, the detector would be broadband

[171]. This solution was impossible in the 70's, but it may become possible soon, with nanotechnology.

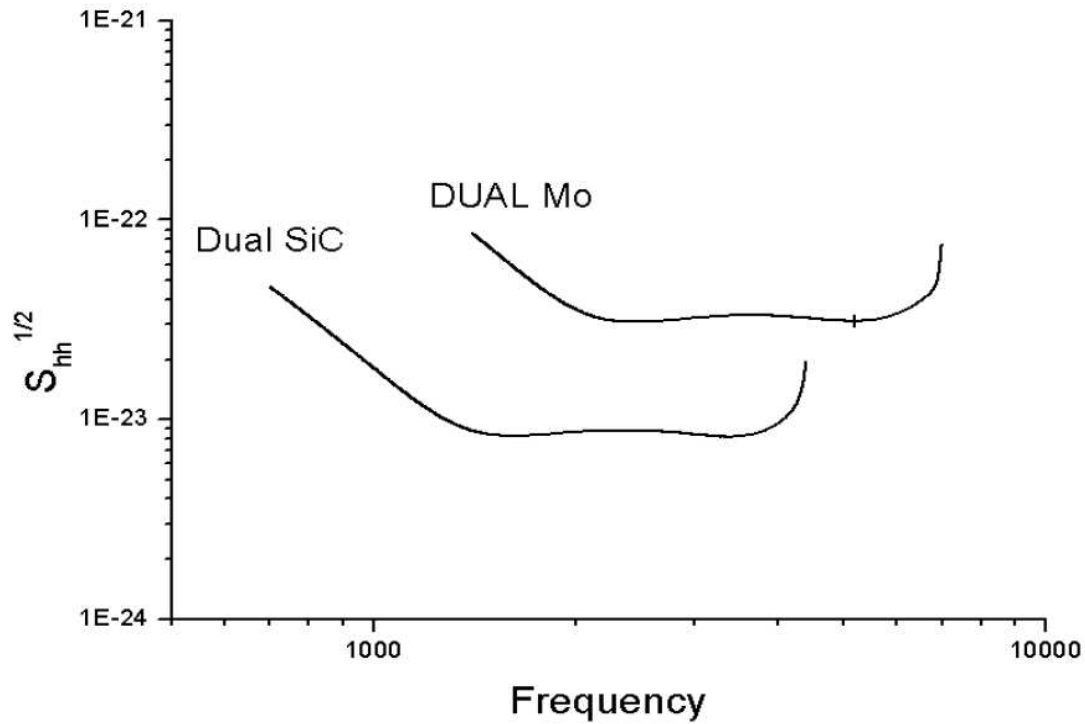

**Figure 7.13.** Sensitivity curves for two dual detectors: Dual SiC detector (inner cylinder radius 0.82 m, outer cylinder internal-external radius 0.83 - 1.44 m, height 3 m, weight 20.5 - 41.7 tons, fundamental quadrupolar modes 3281 Hz and 596 Hz, amplifier noise Sxx = $6 \times 10^{-46}$ m²/Hz, Q/T > $2 \times 10^8$ K$^{-1}$) and Dual Mo dual detector (inner cylinder radius 0.25 m, outer cylinder internal-external radius 0.26 - 0.47 m, height 2.35 m, weight 4.8 + 11.6 tons, fundamental quadrupolar modes 5189 Hz and 1012 Hz, amplifier noise Sxx = $10 \times 10{-46}$ m²/Hz, Q/T > $2 \times 10^8$ K$^{-1}$) (courtesy of the Auriga collaboration, in this case: M. Bonaldi, M. Cerdonio, L. Conti, M. Pinard, G.A. Prodi, L. Taffarello and J.P. Zendri) [169].

Applying this idea, a xylophone of a few spheres was proposed to cover a very large bandwidth using only six non-resonant transducers each [172]. The array of spheres mentioned in previous works [122, 137] required a larger number of spheres than the present proposal, and each of them with a much larger number of resonant-transducers (a set of transducers for each order of quadrupole modes). In figure 7.14 it is shown the sensitivity curve of an array of six spheres (in which the Schenberg is the larger one) using six non-resonant transducers.

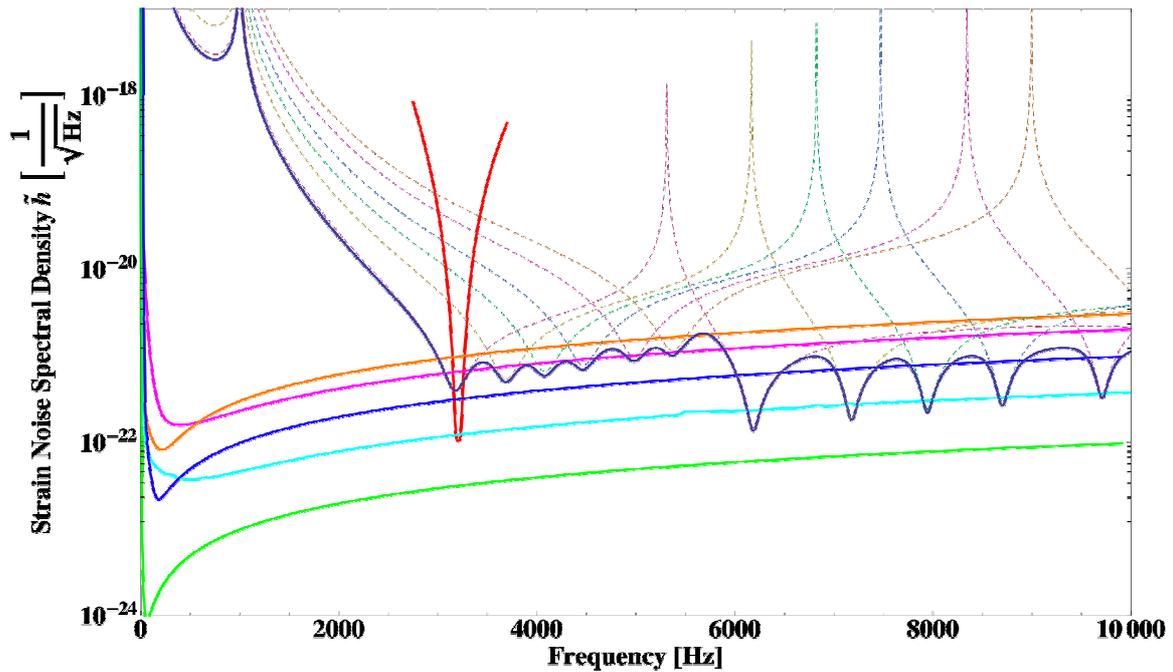

**Figure 7.14.** The sensitivity curve for the Schenberg broadband detector using a nanogap klystron cavity non-resonant transducer. The dashed curves represent each of the 6 spheres we chose for the array (masses: 1150 kg (Schenberg), 744 kg, 547 kg, 414 kg, 301 kg, 239 kg), the lowest frequency being Schenberg. The V-shaped red curve is Schenberg with its usual configuration operating at dilution fridge temperatures (~10 mK). Interferometer curves are also plotted: advanced LIGO (green), LIGO (blue), TAMA300 (pink) and GEO600 (orange). All of these are project curves, not actual data.

Spherical detectors, when fully operational at their designed sensitivities, will be very reliable instruments. The reason for that is because they are six-sensor detectors while the interferometers, as the bars, are only single sensor detectors. A gravitational wave burst signal arrives to the interferometers and to the bars at different times, therefore it is very difficult to confirm a burst signal if it comes close to the noise level. In these cases, a confirmation from the electromagnetic band is necessary, which gives the valuable information of the wave direction and arrival time. On the other hand, being a six-sensor instrument, a spherical antenna has always robust information about a detected signal, which already helps to separate with a high probability a real signal from ordinary noise.

An array of six spheres, when placed close enough together (only a few meters apart) and sampled with the same A/D converter system is an even more reliable system. They will form a coherent system able to provide the correct phase in its various frequency bands, because the wave is a coherent source of energy on the various frequency bands of such a detecting system. Furthermore, detecting a wave with a different physical principle (absorption of the wave energy by the resonant-mass) will certainly contribute to our knowledge about it.

In conclusion, the future of resonant-mass detectors is still promising. New technological improvements will take them to the standard quantum limit and beyond.

# 8. Scientific Outcome of Gravitational Wave Astronomy

The first gravitational wave detection and the regular observation of gravitational waves are certainly among the most important scientific goals and technological challenges for the beginning of this millennium. They will open a new window for the observation of the universe.

Evidently the first impact of gravitational wave detection will be in Gravitational Physics. The direct confirmation of the existence of these waves will be, in itself, a very important scientific milestone. However, it is probably in Astrophysics and Cosmology that the greatest scientific benefit will be found from this "window" of observation, because gravitational waves carry information that cannot be obtained from the detection of electromagnetic waves (radio, infrared, optical, ultraviolet, X- and gamma rays). The information provided by such waves is completely different when compared to that provided by electromagnetic waves. GWs carry detailed information on the coherent bulk motions of matter, such as in collapsing stellar cores or coherent vibrations of space–time curvature as produced, for example, by black holes. On the other hand, electromagnetic waves are usually an incoherent superposition of emissions from individual atoms, molecules and charged particles, and are unable to provide data from the dense core of some astrophysical objects. Gravitational Wave *telesensors* [173] rather than electromagnetic telescopes will "feel" the mass dynamics of a compact object's interior. Therefore, the detection of gravitational waves will open a new window for the observation of the universe.

The speed and polarity of the oncoming GWs will be determined. Later on, as more detections are registered, aspects of gravitation never before observed will become apparent. Then, we will be able to put the General Theory of Relativity and other gravitation theories to the test [174, 175, 144]. General Relativity, for example, predicts that the graviton, the fundamental particle associated with GWs, has zero mass, and spin 2; therefore gravitons travel at the speed of light, carry energy ($E = \hbar \omega$) and momentum ($p = E/c$), and produce the tensorial forces on matter discussed at beginning of this chapter. General Relativity also predicts a very nonlinear gravity at vicinities of high density matter (strong field), a phenomenon which could be confirmed by the detection of black hole coalescing binaries. Other theories of gravitation, either classical or quantum, including braneworld theories, may predict other polarizations, intensities, and speed for the gravitational waves [176, 177, 178, 179, 180].

Because theoretical models estimate that collapsing and bouncing cores of supernovas, as well as corequakes and star collisions, produce large intensities of gravitational radiation in the vicinity of 1-3 kHz, these events are likely to be observed by these resonant-mass detectors, since the pioneering works in the 60's and 70's.

The strain expected to reach Earth coming from one of the burst sources mentioned above is highly uncertain. Thorne [13] gives the expression:

$$h = 2.7 \times 10^{-20} \left[ \frac{\Delta E_{GW}}{M_s c^2} \right]^{1/2} \left[ \frac{1 \text{kHz}}{f} \right]^{1/2} \left[ \frac{10 \text{ Mpc}}{d} \right] \tag{8.1}$$

where $\Delta E_{GW}$ is the energy converted to GWs, $M_S$ is the solar mass, $f$ is the characteristic frequency of the burst, and $d$ its distance from us (~14 Mpc is the distance to the center of the Virgo Cluster of galaxies for a Hubble constant of 70 km/ sec / Mpc). The fraction of energy ($\Delta E_{GW} / M_S c^2$) converted to gravitational radiation depends on the asymmetry of the event. For perfectly symmetrical events this fraction is zero, since there is no variation of the quadrupole moment. A fairly optimistic value for supernova events is an energy fraction around $7 \times 10^{-4}$, (Stark, R. F. and Piran, T. 1986, Proceedings of the 4th Marcel Grossmann Meeting on General Relativity, Elsevier, Amsterdam, page 327) which gives a strain close to $h \sim 10^{-18}$ if the supernova occurs in the center of our galaxy (8.5 kpc away). This strain was in the detection range for the bar detectors since the early 90's [181, 182]. However, no supernova explosion in our galaxy able to produced neutrinos occurred in the past 20 years, as the community of neutrino detectors states. Considering that the bars have today a sensitivity for burst around $h \sim 3 \times 10^{-19}$, they would be able to detect a supernova exploding, with the GW energy factor mentioned above, as far as 25 kpc (so, in our galaxy). Therefore, the time for this to happen may come soon.

There is a host of possible astrophysical sources of gravitational waves: namely, supernovae, the collapse of a star or star cluster to form a black hole, inspiral and coalescence of compact binaries, MACHOs as primordial black holes (PBHs), quark stars, boson stars, neutron star modes, the fall of stars and black holes into supermassive black holes, rotating neutron stars, ordinary binary stars, relics of the big bang, vibration or collision of monopoles, cosmic strings and cosmic bubbles, amongst others [13, 183, 184, 185, 186, 187, 188, 189]. From the theoretical point of view there has been a great effort to study which are the most promising sources of gravitational waves to be detected.

In particular, the waveforms, the characteristic frequencies and the number of sources per year that one expects to observe are questions that have been addressed [184, 186, 190]. In a few years, starting from the observations (waveforms, amplitudes, polarizations, etc.), it will really be possible to understand how gravitational wave emission is generated by astrophysical sources.

It is quite probable that the Universe is pervaded by a background of gravitational waves, because of the fact that they are produced by a large variety of astrophysical sources and cosmological phenomena. A variety of binary stars (ordinary, compact or combinations of them), Population III stars, phase transitions in the early Universe and cosmic strings and a variety of binary stars are examples of sources that could produce such a putative background of gravitational waves [13, 191, 192, 193, 194, 186, 195, 196, 197, 198].

With a network of sufficient sensitive gravitational wave detectors important information could be obtained about [13]:
- the existence of black holes;
- the position of nearby neutron stars and black holes, and their ratio (number of neutron stars) / (number of black holes);

- the formation rate of compact stars in our galaxy and neighboring galaxies and the fraction of supernovas that are not visible;
- the explosion mechanism, including the dynamics (rotation, radial movement, and timescale of collapse and rebound) of the nucleus' collapse, as well as the masses and angular momenta of supernovae occurring in our galaxy and in neighboring galaxies;
- the masses and viscosities of neutron stars (including their upper limit), and the masses of black holes (including their lower limit);
- the equation of state of neutron star and quark (or strange) matter if it exists;
- the sources of gamma-ray bursts, if they are associated to compact binaries [199];
- the hypothesis of occurrence of hydrodynamic instability in neutron stars by gain of angular momentum due to accretion matter in binary systems [200];
- the amount of baryonic dark matter in the local universe;
- the determination of the Hubble constant with reasonable precision, checking the values of current observations [201], which also includes the possible accelerated expansion of the Universe;
- the cosmological structure of the local Universe.

The gravitational wave astronomy may be very important for testing other theories of gravitation such as brane-world theories, the existence of extra dimensions [202], the holographic principle, and cosmological theories which try to explain the existence of dark matter and dark energy [203, 204].

## 9. Summary

The first experiment to detect gravitational waves was proposed and performed over 40 years ago by Weber using a resonant-mass detector with a bar shape. Around January 1965 the first high-frequency detector, designed and constructed by Weber, went into operation. Soon he started claiming coincidences between two operational detectors 1000 km apart. During the seventies and the years that followed there were unsuccessful efforts worldwide to repeat Weber's claims even with higher sensitivity. To date there is no conclusive confirmation of such events, and theoretical estimates of local fluxes carried out so far also reinforce this position. However, Weber's pioneering work was decisive for the initial growth of the gravitational wave community. Thanks to Weber's results more than a dozen gravitational wave detector groups worldwide were formed, and the basis of gravitational wave detection was established.

Today resonant-mass bar detectors have achieved a superb stable operation. Some of them have very high sensitivity and duty cycles above 95%. Simultaneous operation of four bar detectors for long periods of time is a technological possibility nowadays. Their sensitivity can also be improved if enough financial support is given. The standard quantum limit can be surpassed if state-of-art parametric transducers are used. New resonant-mass projects have been pursued over the last decade: spherical antennae promise a leap in sensitivity and capability. Direction and polarization will be determined by the use of a single spherical detector. Simultaneous operation of an array of such detectors may become a reality soon. If that is not enough, dual resonant-mass antennae will also provide resonant-mass detectors with a much larger bandwidth at very low cost, surpassing the xylophone (an array of detectors each of which tuned to a specific frequency) proposal. A xylophone of spheres equipped with non-resonant transducers, on the other hand, may become the best solution for wideband gravitational wave detection, capable to find the wave's direction and polarization with a single site instrument. A cubic array of spheres could even make directional detection by suitable computational analysis of the spheres' outputs.

The quest for gravitational wave detection has been one of the toughest technological challenges ever faced by experimental physicists and engineers. Despite the null results to date, after four decades of research, the community involved in this area is continuously growing. One of the main reasons for this is because the first gravitational wave detection and the regular observation of gravitational waves are among the most important scientific goals for the beginning of this millennium. They will test one of the foundations of physics, Einstein's theory of general relativity, and will open a new window for the observation of the universe, which certainly will cause a revolution in our knowledge of physics and astrophysics.

## Acknowledgments

Many thanks to Guido Pizzella, Bill Hamilton, Massimo Cerdonio, Warren Johnson, Eugenio Coccia, David Blair, Kimio Tsubono, Walter Winkler, Giorgio Papini, Viviana Fafone, Jean-Pierre Zendri, Giovanni Prodi, Michele Bonaldi, Gianfranco Giordano, Ik Siong Heng, and many other people for providing me with valuable information about resonant-mass detectors (history, data, pictures, and graphs). I also would like to thank the IGEC2 collaboration (Auriga-ROG-Allegro) for giving me the permission to use pictures, graphs and other data about the present status of the bar detectors. I want to thank all members and collaborators of the Graviton project, in special Nei Oliveira Jr.; without them the Schenberg detector would not be a reality today. Finally, I want to thank José Antônio Pacheco for the opportunity he has given to the resonant-mass community. This work has been supported by FAPESP (under grant No.1998/13468-9 and 2006/56041-3), CNPq (under grant No. 306467/03-8), CAPES and MCT/INPE.